\documentstyle[preprint,eqsecnum,amsfonts,aps]{revtex}

\newcommand{\sous}[2]{\stackrel{\phantom{(n,p)}}{#2} \! \! \! \! \! \!
			\! \! \! 
			{{} \atop \scriptstyle #1} {}}

\begin{document}
\draft
\title{Lorentzian regularization and the problem \\
of point-like particles in general relativity}
\author{Luc Blanchet}
\address{D\'epartement d'Astrophysique Relativiste et de
Cosmologie, \\ 
Centre National de la Recherche Scientifique (UMR 8629),\\
Observatoire de Paris, 92195 Meudon Cedex, France,\\ 
and
Department of Earth and Space Science,\\
Graduate School of Science, Osaka University,\\
Toyonaka, Osaka 560--0043, Japan}

\author{Guillaume Faye}
\address{D\'epartement d'Astrophysique Relativiste et de Cosmologie,\\
Centre National de la Recherche Scientifique (UMR 8629),\\
Observatoire de Paris, 92195 Meudon Cedex, France}

\date{\today}
\maketitle
\widetext
\begin{abstract}
The two purposes of the paper are (1) to present a regularization of
the self-field of point-like particles, based on Hadamard's concept of
``partie finie'', that permits in principle to maintain the Lorentz
covariance of a relativistic field theory, (2) to use this
regularization for defining a model of stress-energy tensor that
describes point-particles in post-Newtonian expansions (e.g. 3PN) of
general relativity. We consider specifically the case of a system of
two point-particles. We first perform a Lorentz transformation of the
system's variables which carries one of the particles to its rest
frame, next implement the Hadamard regularization within that frame,
and finally come back to the original variables with the help of the
inverse Lorentz transformation. The Lorentzian regularization is
defined in this way up to any order in the relativistic parameter
$1/c^2$. Following a previous work of ours, we then construct the
delta-pseudo-functions associated with this regularization.  Using an
action principle, we derive the stress-energy tensor, made of
delta-pseudo-functions, of point-like particles. The equations of
motion take the same form as the geodesic equations of test particles
on a fixed background, but the role of the background is now played by
the regularized metric.
\end{abstract}


\narrowtext

\section{Introduction}\label{I}

In recent years, the problem of the dynamics of gravitationally
interacting compact objects in general relativity has received a lot
of attention. This is due in part to the interest of the theoretical
problem in its own, and in part to the ongoing development of
laser-interferometric detectors for observing gravitational
radiation. In the absence of an exact solution of the problem, one has
recourse to successive post-Newtonian approximations (formal
expansions in powers of $1/c$). Within such approximations, it makes
sense to model the compact objects with some ``point-like particles'',
exactly as we do in a standard way within the Newtonian
theory. However, the self-field of point-particles is infinite at the
very location of a particle, and thus must be somehow regularized. The
regularization is quite straightforward in the Newtonian theory, but
it becomes non-trivial when going to high post-Newtonian
approximations. Dealing with this problem, the present authors
\cite{BFreg} developed a method for regularizing the infinite
self-field of point-particles, which is based on the concept of
``partie finie'', in the sense of Hadamard \cite{Hadamard,Schwartz},
of a singular function at the place of one of its singular points (see
e.g. \cite{EstrK85,EstrK89,Sellier,Jones96} for entries to the
mathematical literature). We know that the Hadamard regularization
yields the correct result for the equations of motion of two particles
up to the so-called second and half post-Newtonian (2.5PN)
approximation, corresponding to the order $1/c^5$ beyond the Newtonian
acceleration. Indeed, the problem has been completely solved at that
order
\cite{LD17,EIH,O74a,O74b,BeDD81,DD81a,D83a,S85,S86,Kop85,GKop86,BFP98};
notably some derivations make use of this regularization
(e.g. \cite{BeDD81,BFP98}). In the present state of the art, we are
concerned with the 3PN (or $1/c^6$) approximation
\cite{JaraS98,JaraS99,BF00,DJS00,BFeom,ABF01}. In fact, starting at this
high post-Newtonian order, the regularization may become physically
incomplete because of the appearance of an undetermined coefficient in
the equations of motion \cite{JaraS98,JaraS99,BF00,DJS00,BFeom,ABF01}.

The Hadamard regularization, investigated in \cite{BFreg}, is
performed in a three-dimensional Euclidean space with generic point
${\bf x}\in {\mathbb R}^3$, which is viewed as the spatial
hypersurface labelled by $t=$const in a global coordinate system
$\{t,{\bf x}\}$ covering the whole space-time. In particular, the
regularization involves a spatial average, performed at $t=$const,
over the direction of approach to the singularity. As such a
regularization makes use of a preferred spatial hypersurface
$t=$const, it is clearly incompatible with the framework of special or
general relativity, which embodies a global Lorentz (or Poincar\'e)
frame invariance. Notably, we expect that the post-Newtonian equations
of motion of point-like particles in harmonic coordinates (which we
recall preserve the global Lorentz invariance) should exhibit at some
stage a violation of the Lorentz invariance due to the latter
regularization. The fact is that the breakdown of the Lorentz
invariance due to the regularization occurs only at the very high 3PN
approximation. Untill the 2.5PN order, it is sufficient to regularize
within a preferred slice $t=$const of the harmonic coordinate system
to obtain some Lorentz-invariant equations of motion \cite{BFP98}.

The first purpose of this paper is to define a regularization {\it \`a
la} Hadamard \cite{Hadamard,Schwartz} that is compatible with the
Lorentz structure of a relativistic field theory. This completes the
definition, proposed in \cite{BFreg}, of a specific version of the
Hadamard regularization (based notably on a particular class of
pseudo-functions). To achieve this purpose, we shall simply perform
the standard Hadamard regularization within the hypersurface that is
geometrically orthogonal, in the sense of the Minkowski metric, to the
four-velocity of the particle. In separate papers \cite{BFeom,ABF01},
we apply the latter ``Lorentzian'' regularization (together with the
distributional derivatives introduced in \cite{BFreg}), to the
computation of the binary equations of motion at the 3PN order in
harmonic coordinates, and find that, indeed, it permits the preserving
of their Lorentz invariance (in some case at the price of adjusting
some parameter). A different approach to the problem of incorporating
the Lorentz invariance in the 3PN equations of motion consists of
deriving a generic regularized dynamics, within the ADM-Hamiltonian
formalism of general relativity, involving an arbitrary regularization
parameter, and to determine this parameter uniquely by requiring the
Lorentz invariance \cite{DJS00}. (See Section 2 in \cite{ABF01} for a
discussion on our point-mass regularization and its relation to
\cite{DJS00}.)

All-over the paper, we assume the existence of a preferred Minkowski
metric, as selected for instance by the condition of harmonic
coordinates in general relativity, with respect to which the
trajectories of particles are represented by accelerated world lines
like in special relativity. Most of our investigation is valid not
only in the case of the gravitational field but also for any
Lorentz-tensor field propagating on the Minkowski
background. Furthermore, we shall define the Lorentzian regularization
in a sense of formal expansion series in $1/c^2$; so that, all the
formulas in the paper will be given by some infinite series of
relativistic corrections when $c$ tends toward infinity. This is all
what we need for the derivation of the equations of motion to the 3PN
order \cite{BFeom,ABF01}. 

Since we are interested in the application to the motion of two
particles, we shall define the regularization around one of the
particles (say particle 1), and shall consider that its acceleration
is purely due to particle 2. [However our definitions could be
generalized to a system of N particles.] Notice that the particle 2
enters this regularization scheme through the Lorentz transformation
of its own variables to the rest frame of particle 1, and the
replacement of the acceleration of 1 in terms of the equations of the
binary motion. In general, working at some given relativistic order,
we shall need to know the equations of motion up to a lower order
only, therefore giving us the possiblity of an iterative process. In
this paper, we always assume that we know the relevant equations of
motion at this order, and that these are Lorentz-invariant.

Our second purpose is to derive an expression, compatible with the
latter regularization, for the stress-energy tensor of point-like
particles in post-Newtonian expansions of general relativity. Thanks
to this regularization, we are able to give a sense to the value of
the metric coefficients at the very location of the particle. Our
basic assumption is that the matter action is the same as for {\it
test} particles moving on a prescribed background gravitational field,
except that the metric at the location of the particles is replaced by
its regularized value in the sense of the (Lorentzian)
regularization. From this assumption, we prove that the Dirac measure
in the stress-energy tensor must be replaced by a certain generalized
function defined by means of the Hadamard prescription. In the case of
two particles (the generalization to $N$ particles is immediate), we
obtain

\begin{equation}\label{i1}
T^{\mu\nu}_{\rm particle}={m_1 c ~\!v_1^\mu v_1^\nu \over
\sqrt{-[g_{\rho\sigma}]_{\textstyle {}_1} v_1^\rho v_1^\sigma}}
~\!{\rm Pf}\left({\Delta ({\bf x}-{\bf y}_1)\over \sqrt{g(t,{\bf
x})}}\right) + 1\leftrightarrow 2 \;,
\end{equation}
where $m_1$ is the mass of the particle 1, and $v_1^\mu=(c,{\bf v}_1)$
its coordinate velocity, i.e. ${\bf v}_1=d{\bf y}_1/dt$, ${\bf
y}_1={\bf y}_1(t)$ being the trajectory parametrized by the coordinate
time $t$ (the symbol $1\leftrightarrow 2$ denotes the same expression
but corresponding to the second particle). The notation
$[g_{\rho\sigma}]_{\textstyle {}_1}$ means that the metric
$g_{\rho\sigma}(t,{\bf x})$ is to be computed at the point ${\bf
x}={\bf y}_1(t)$ following the regularization (of course
$[g_{\rho\sigma}]_{\textstyle {}_1}$ depends on the positions and
velocities of both particles 1 and 2). Note that the first factor in
(\ref{i1}) is a mere function of time $t$. The second factor ${\rm
Pf}\left({\Delta ({\bf x}-{\bf y}_1)\over \sqrt{g}}\right)$ is made of
a special type of partie finie delta-pseudo-function associated with
the regularization (following the definition given in
\cite{BFreg}). It involves (minus) the determinant of the metric
$g_{\rho\sigma}$, namely $g$, evaluated at the point $(t,{\bf x})$,
and a generalization ${\rm Pf}\Delta({\bf x}-{\bf y}_1)$ of the Dirac
function defined in such a way that its action on a singular function
yields the value of the function at the singular point in the sense of
the regularization. Among the rules for handling the
delta-pseudo-functions, we are allowed to write ${\rm Pf}\left({\Delta
({\bf x}-{\bf y}_1)\over \sqrt{g}}\right)=\case{1}{\sqrt{g}}{\rm
Pf}\Delta ({\bf x}-{\bf y}_1)$, whereas it is strictly forbidden to
replace the latter quantity by $[\case{1}{\sqrt{g}}]_{\textstyle
{}_1}{\rm Pf}\Delta ({\bf x}-{\bf y}_1)$.

The stress-energy tensor (\ref{i1}) takes the same form as the one of
test particles moving in a fixed background, but with the role of the
background played by the regularized metric generated by the
bodies. In particular, the equations of motion obtained from the
covariant conservation of that tensor ($\nabla_\nu T^{\mu\nu}_{\rm
particle}=0$), take the same form as the ``geodesic equations'', when
considered with respect to the regularized metric. Our definition of
the stress-energy tensor (\ref{i1}) constitutes a proposal, that we
have found to be the most natural in the problem of the equations of
binary motion at the 3PN order \cite{BFeom,ABF01}, but that we have
not proved to be generally valid to higher post-Newtonian orders (nor
of course when considered outside a framework of post-Newtonian
expansions). The tensor (\ref{i1}) appears to be a good candidate for
the characterization of point-like particles in post-Newtonian
expansions of general relativity.

The plan of this paper is the following. In Section \ref{II}, we
recall from \cite{BFreg} the material needed in the subsequent parts
concerning the Hadamard regularization and the associated
pseudo-functions. In Section \ref{III}, we investigate the formulas,
needed to regularize, for the Lorentz transformation of some field
point as well as two source points, and we define the new
regularization around one of the particles as taking place within the
instantaneous spatial hypersurface of the particle. In Section
\ref{IV}, we give the formulas for this regularization at the level of
the first relativistic correction $1/c^2$. Finally, in Section
\ref{V}, we derive from an action principle our model of stress-energy
tensor of point-like particles; the covariant conservation of this
tensor leads to the equations of motion.

\section{Hadamard regularization}\label{II} 

To make the present paper self-contained, we shall review in this
section the classic notions of the Hadamard regularization of singular
functions and divergent integrals \cite{Hadamard,Schwartz}, as well as
the construction, by Blanchet and Faye \cite{BFreg}, of a set of
pseudo-functions associated with it. We follow closely the
investigation of our previous paper \cite{BFreg} and employ most of
its notation. A coordinate system $\{t,{\bf x}\}$ being given on
space-time (for instance the harmonic coordinates used in Section IV),
we consider some functions $F({\bf x})$ defined on the spatial slice
$t=$const, where ${\bf x}\in {\mathbb R}^3$ denotes the position in
the slice.  We say that the function $F({\bf x})$ belongs to the class
${\cal F}$ if and only if $F$ is a smooth function on ${\mathbb R}^3$
except at two isolated points ${\bf y}_1$ and ${\bf y}_2$, and admits
around each of these points the following power-like singular
expansions. Denoting by $r_1=|{\bf x}-{\bf y}_1|$ the spatial distance
to the point 1, and by ${\bf n}_1=({\bf x}-{\bf y}_1)/r_1$ the spatial
direction of approach to 1, we assume that, for any $N\in {\mathbb
N}$,

\begin{equation}\label{1}
F({\bf x})=\sum_{a_0\leq a\leq N} r_1^a \!\!\sous{1}{f}_{a}({\bf n}_1)+o(r_1^N)\;.
\end{equation}
The coefficients ${}_1f_a$ of the various powers of $r_1$ are smooth
functions of the unit vector ${\bf n}_1$, and the remainder tends to
zero strictly more rapidly than $r_1^N$ when $r_1\to 0$. The powers
$a$ of $r_1$ in that expansion are assumed to be real, $a\in {\mathbb
R}$, to range in discrete steps, i.e. $a\in (a_i)_{i\in {\mathbb N}}$,
and to be bounded from below, i.e. $a_0\leq a$ for some $a_0\in
{\mathbb R}$. Similarly, we assume the same type of expansion around
the point 2,

\begin{equation}\label{2}
\forall N\in {\mathbb N}\;,\quad F({\bf x})=\sum_{b_0\leq b\leq N} 
r_2^b \!\!\sous{2}{f}_{b}({\bf n}_2)+o(r_2^N)\;,
\end{equation}
where $r_2=|{\bf x}-{\bf y}_2|$ and ${\bf n}_2=({\bf x}-{\bf
y}_2)/r_2$.  Thus, to each function $F$ in the class ${\cal F}$ are
associated two discrete families of indices $a$ and $b$, and two
corresponding families of coefficients ${}_1f_a({\bf n}_1)$ and
${}_2f_b({\bf n}_2)$, all of them depending on $F$. We shall refer to
the coefficients ${}_1f_a$ for which $a<0$ (and similarly to ${}_2f_b$
when $b<0$) as the {\it singular} coefficients of $F$ in the expansion
when $r_1\to 0$. Since $a\geq a_0(F)$ and $b\geq b_0(F)$, the number
of singular coefficients of $F$ is always finite.

The so-called ``partie finie'' in the sense of Hadamard
\cite{Hadamard,Schwartz} of the singular function $F$ at the location
of the singular point ${\bf y}_1$ is equal to the angular average, say
${}_1{\hat f}_0$, of the zeroth-order coefficient, ${}_1f_0({\bf
n}_1)$, in the expansion of the function when $r_1\to 0$ we assumed in
(\ref{1}); namely

\begin{equation}\label{3}
(F)_{\textstyle {}_1}= \sous{1}{\hat f}_0 \equiv \int {d\Omega_1\over 4\pi} 
\!\!\sous{1}{f}_0({\bf n}_1)\;,
\end{equation}
where $d\Omega_1= d\Omega ({\bf n}_1)$ denotes the solid angle element
of origin ${\bf y}_1$ and direction ${\bf n}_1$; the latter angular
integration is performed within the coordinate hypersurface
$t=$const. A crucial property of the Hadamard partie finie is its
``non-distributivity'' with respect to the multiplication, in the
sense that

\begin{equation}\label{4}
(FG)_{\textstyle {}_1} \not= (F)_{\textstyle {}_1} (G)_{\textstyle {}_1}
\end{equation}
in general. When applied to the gradient $\partial_iF$ of a function
$F\in {\cal F}$, the definition (\ref{3}) yields a useful formula
which permits one to compute rapidly the partie finie of complicated
expressions involving gradients:

\begin{equation}\label{5}
(\partial_iF)_{\textstyle {}_1}=3\biggl({n_1^i\over r_1}F\biggr)_1\;.
\end{equation}

Closely related to the concept of partie finie of a singular function
is the definition of the partie finie (${\rm Pf}$) of the divergent
integral $\int d^3{\bf x}~F$. All-over this paper, we assume that the
functions decrease fast enough at infinity (when $|{\bf x}|\to
+\infty$) so that the possible divergencies of integrals come only
from the bounds located at the two singular points 1 and 2. The
``partie-finie integral'' reads \cite{Hadamard,Schwartz} as

\begin{eqnarray}\label{6}
{\rm Pf}_{s_1,s_2}\int d^3{\bf x}~ F &=&~\lim_{s\to
0}~\Biggl\{\int_{{\mathbb R}^3\setminus {\sc B}_1(s)\cup {\sc
B}_2(s)}d^3{\bf x}~ F\nonumber\\ &&\qquad +~4\pi\sum_{a+3<
0}{s^{a+3}\over a+3} \left({F\over r_1^a}\right)_1+ 4 \pi
\ln\left({s\over s_1}\right) \left(r_1^3 F\right)_1 +1\leftrightarrow
2\Biggr\}\;.
\end{eqnarray}
The integral in the right side extends over ${\mathbb R}^3$ deprived
from two closed spherical balls ${\sc B}_1(s)$ and ${\sc B}_2(s)$ of
radius $s$ centered on the two singularities [thus ${\sc B}_1(s)$ and
${\sc B}_2(s)$ are defined by $r_1\leq s$ and $r_2\leq s$]. The other
terms, which are defined by means of the partie finie in the sense of
(\ref{3}), are chosen in such a way that the limit $s\to 0$
exists. The notation $1\leftrightarrow 2$ indicates the same terms as
the two previous ones but corresponding to the other point. The
summation index $a$ satisfies $a_0\leq a <-3$ (in particular the sum
is always finite). Notice the two arbitrary constants $s_1$ and $s_2$
which are introduced in order to adimensionalize the arguments of the
logarithms in (\ref{6}); the partie finie owns an ambiguity through
these constants (hence the notation ${\rm Pf}_{s_1,s_2}$).  The close
connection between the partie finie of a singular function (\ref{3})
and that of a divergent integral (\ref{6}) is most easily seen from
the fact that \cite{BFreg}

\begin{equation}\label{8}
{\rm Pf}\int d^3{\bf x}~ \partial_i F = -4 \pi 
(n_1^i r_1^2 F)_{\textstyle {}_1} +1\leftrightarrow 2\;.
\end{equation}
Unlike in the case of continuous functions, the (partie-finie)
integral of a gradient is non-zero in general, and equal to the sum of
the parties finies, in the sense of (\ref{3}), of the surface
integrals surrounding the singularities, in the limit where the
surface areas tend to zero. This fact motivated the introduction and
study in \cite{BFreg} of a new derivative operator acting on ${\cal
F}$, satisfying a property of ``integration by parts'' implying that
the integral of any gradient is always zero. This operator generalizes
for the class of functions ${\cal F}$ the standard distributional
derivative of Schwartz \cite{Schwartz}.

Let us associate to any $F\in {\cal F}$ a pseudo-function denoted
${\rm Pf} F$ and defined to be the following linear form acting on the
class ${\cal F}$~:

\begin{equation}\label{9}
\forall G \in {\cal F}\;,\quad < {\rm Pf} F, G > ={\rm Pf} \int d^3{\bf x}~ F G\;,
\end{equation}
where the right side is a partie-finie integral in the sense of
(\ref{6}); we use a duality bracket to denote the result of the action
of the pseudo-function ${\rm Pf}F$ on $G$. A fundamental definition
adopted in \cite{BFreg}, and motivated by the application to Physics,
concerns the product of two pseudo-functions, or of a function and a
pseudo-function, which is the ``{\it ordinary}'' pointwise product in
the sense that

\begin{equation}\label{10}
{\rm Pf}F~\!.~\!{\rm Pf}G=F~\!.~\!{\rm Pf}G=G~\!.~\!{\rm Pf}F={\rm Pf}(FG)\;.
\end{equation} 
Thus, for instance,

\begin{equation}\label{11}
< {\rm Pf}F~\!.~\!{\rm Pf}G, H > ={\rm Pf} \int d^3{\bf x}~ F G H\;.
\end{equation}
The product (\ref{10}) chosen in \cite{BFreg} dictates most of the
subsequent properties of the pseudo-functions, as well as their
generalized distributional derivatives. (Refer to
\cite{Gelfand,Jones82,Kanwal83} for mathematical treatises on
generalized functions and distributions.) In particular, the
derivatives do not in general satisfy the Leibniz rule for the
derivation of the product, although they satisfy it in an ``integrated
sense'', according to the rule of integration by parts.
 
The Riesz \cite{Riesz} delta-function, given for $\varepsilon >0$ by
${}_\varepsilon\delta({\bf x})=\case{\varepsilon
(1-\varepsilon)}{4\pi}~\!|{\bf x}|^{\varepsilon-3}$, tends, in the
usual sense of distribution theory, towards the Dirac measure when
$\varepsilon\to 0$. When considered with respect to the singular point
${\bf y}_1$, the Riesz delta-function allows us to define a useful
element of our class,

\begin{equation}\label{12}
{}_\varepsilon\delta_1({\bf x})\equiv {}_\varepsilon\delta({\bf
x}-{\bf y}_1)=\case{\varepsilon
(1-\varepsilon)}{4\pi}~\!r_1^{\varepsilon-3}~\in~{\cal F}\;.
\end{equation}
Therefore it is possible to associate to ${}_\varepsilon\delta_1$ (for
any $\varepsilon>0$) the pseudo-function ${\rm
Pf}{}_\varepsilon\delta_1$ following the prescription
(\ref{9}). Applying the limit $\varepsilon\to 0$, we obtain
\cite{BFreg}

\begin{equation}\label{13}
\lim_{\varepsilon\to 0}~\!<{\rm Pf}{}_\varepsilon\delta_1,F>\equiv 
\lim_{\varepsilon\to 0}~{\rm Pf}\int d^3{\bf x}~ 
{}_\varepsilon\delta_1 F = (F)_{\textstyle {}_1}\;,
\end{equation}
where the value of $F$ at the point 1 in the right side is defined by
the prescription (\ref{3}). This motivates us for introducing a new
pseudo-function, we shall call the delta-pseudo-function ${\rm
Pf}\delta_1$, as the formal limit of the pseudo-functions ${\rm
Pf}{}_\varepsilon\delta_1$ when $\varepsilon\to 0$. By definition,

\begin{equation}\label{14}
\forall F\in {\cal F},\quad
<{\rm Pf}\delta_1,F>=(F)_{\textstyle {}_1} \;.
\end{equation}
Clearly, the delta-pseudo-function ${\rm Pf}\delta_1$ generalizes the
notion of Dirac distribution $\delta_1\equiv \delta ({\bf x}-{\bf
y}_1)$ to the case where the ``test'' functions are singular and
belong to the class ${\cal F}$. Extending the definition of the
product (\ref{10}) to include the delta-pseudo-function we pose

\begin{equation}\label{15}
{\rm Pf}F~\!.~\!{\rm Pf}\delta_1=F~\!.~\!{\rm Pf}\delta_1={\rm Pf}(F\delta_1)\;,
\end{equation}
as well as, for instance,

\begin{equation}\label{15'}
{\rm Pf}(F\delta_1)~\!.~\!{\rm Pf}G={\rm Pf}(F\delta_1)~\!.~\!G={\rm Pf}(FG\delta_1)\;.
\end{equation}
The new object ${\rm Pf}(F\delta_1)$ in (\ref{15})-(\ref{15'}) has no
equivalent in distribution theory; it satisfies

\begin{equation}\label{16}
\forall G\in {\cal F}\;,\quad
<{\rm Pf}(F\delta_1),G>=(FG)_{\textstyle {}_1}\;.
\end{equation}
We notice for future reference that a consequence of the
``non-distributivity'' of the Hadamard partie finie [see (\ref{4})] is
that

\begin{equation}\label{16'}
{\rm Pf}(F\delta_1)\not= (F)_{\textstyle {}_1}~{\rm Pf}\delta_1 \;.
\end{equation}
We are not allowed to replace a singular function that appears in
factor of the delta-pseudo-function at the point 1 by its regularized
value at that point.

The derivative of the delta-pseudo-function ${\rm Pf}\delta_1$ was
constructed in \cite{BFreg}. As it turns out, it takes the form of an
``ordinary'' derivative~: $\partial_i({\rm Pf}\delta_1)={\rm
Pf}(\partial_i\delta_1)$; due to the presence of the
delta-pseudo-function, there are no distributional terms associated
with it. We have simply (from the rule of integration by parts),

\begin{equation}\label{16a}
\forall F\in{\cal F}\;,\qquad <\partial_i({\rm Pf}\delta_1),F>
=-<{\rm Pf}\delta_1,\partial_iF>=-(\partial_iF)_{\textstyle {}_1}\;.
\end{equation}
The differentiation of the more complicated object ${\rm
Pf}(F\delta_1)$ proceeds in the same way:

\begin{equation}\label{16b}
\forall G\in{\cal F}\;,\qquad <\partial_i[{\rm Pf}(F\delta_1)],G>
=-<{\rm Pf}(F\delta_1),\partial_iG>=-(F\partial_iG)_{\textstyle {}_1}\;.
\end{equation}
Note that, as a consequence of the identity (\ref{5}), we can write 
for the intrinsic form of this object:

\begin{equation}\label{16c}
\partial_i[{\rm Pf}(F\delta_1)]=
{\rm Pf}\Biggl[r_1^3\partial_i\left({F\over r_1^3}\right)\delta_1\Biggr]\;.
\end{equation}
Because the derivative of the delta-pseudo-function is equal to the
ordinary one, the Leibniz rule for the derivative of a product happens
to still hold. For instance, in the case of the product of ${\rm
Pf}(F\delta_1)$ with some pseudo-function ${\rm Pf}G$, we have

\begin{equation}\label{16d}
\partial_i[{\rm Pf}(F\delta_1)~\!\!.~\!{\rm Pf}G]=
\partial_i[{\rm Pf}(F\delta_1)]~\!\!.~\!{\rm Pf}G+{\rm Pf}(F\delta_1)
~\!\!.~\!\partial_i({\rm Pf}G)\;.
\end{equation}
The proof uses the combination of (\ref{15'}) and (\ref{16b}).

\section{Lorentzian regularization}\label{III}

To define a Lorentzian regularization {\it \`a la} Hadamard (based on
the investigation of \cite{BFreg} and on Section \ref{II}), we now
need to specify in a precise way the dependence of a function $F({\bf
x})$ in the class ${\cal F}$ on the ``source'' variables at the
coordinate time $t$ of a global frame $\{{\bf x},t\}$. We assume (as
everywhere else in this paper) that we are working at some given
finite order in a relativistic or post-Newtonian approximation. Up to
a given order, we can choose as the source variables the two
trajectories ${\bf y}_1(t)$ and ${\bf y}_2(t)$ in the frame $\{{\bf
x},t\}$, and the two coordinate velocities ${\bf v}_1(t)=d{\bf
y}_1/dt$ and ${\bf v}_2(t)=d{\bf y}_2/dt$ (the trajectories of the
particles are time-like world lines in Minkowski space-time). Indeed,
it is legitimate to assume only the latter source variables because,
up to a given post-Newtonian order, we can order-reduce the
accelerations and all derivatives of accelerations by means of the
equations of motion of the particles up to the appropriate accuracy
(in general the precision of the equations of motion needed for this
order-reduction is one order less than the given post-Newtonian order
at which we are performing a calculation). Of course, we are assuming
that these equations of motion are known (they are known presently to
the 2.5PN order \cite{DD81a,D83a,BFP98}, and the general motivation of
this work is to get them up to the 3PN order
\cite{BF00,BFeom,ABF01}. Thus, we assume that the function $F\in {\cal F}$
really reads

\begin{equation}\label{17}
F({\bf x},t)=F[{\bf x};{\bf y}_1(t),{\bf y}_2(t);{\bf v}_1(t),{\bf v}_2(t)]\;.
\end{equation}
We denote with the same letter $F$, by a slight abuse of notation, the
function of the field point $({\bf x},t)$ and the functional of the
field point and source variables in the right-hand-side.  For
definiteness, we assume that the two trajectories are smooth functions
of time, i.e. ${\bf y}_1$, ${\bf y}_2\in C^\infty({\mathbb R}^3)$, and
that $F$ is a smooth functional of the two velocities ${\bf v}_1$,
${\bf v}_2$ (see also Section IX of \cite{BFreg} for details about our
assumptions). By (\ref{17}), we mean that the dependence of $F$ on the
coordinate time $t$ is through (and only through) the two
instantaneous trajectories ${\bf y}_1$, ${\bf y}_2$ and velocities
${\bf v}_1$, ${\bf v}_2$. Note also that it is implicitely assumed
with our notation (\ref{17}) that the function $F$ depends {\it
locally} on time $t$ (no dependence over the trajectories and
velocities at some time earlier than $t$ for instance).  Furthermore,
very often in applications, we shall find that the dependence of $F$
on the spatial position ${\bf x}$ appears only via the two spatial
distances to the source points, ${\bf r}_1(t)={\bf x}-{\bf y}_1(t)$
and ${\bf r}_2(t)={\bf x}-{\bf y}_2(t)$. In this paper, we shall
generally suppose, in order to simplify the presentation, that this is
the case; namely, the function $F$, as a functional of the source
variables, is

\begin{equation}\label{18}
F({\bf x},t)=F[{\bf r}_1(t),{\bf r}_2(t);{\bf v}_1(t),{\bf v}_2(t)]\;.
\end{equation}
The hypothesis (\ref{18}) does not constitute a very severe
restriction. The extension to the more general case (\ref{17}) is
generally straightforward; moreover, (\ref{18}) is always verified in
the problem of the post-Newtonian equations of motion of binary
systems. In this section, we shall define the Lorentzian regularized
value of the function $F$ at the location of the singularity 1, by
contrast to the non-invariant regularized value defined by (\ref{3})
within the ``global'' coordinate hypersurface $t=$const. We shall
denote by $[F]_{\textstyle {}_1}$ the new Lorentzian regularization of
$F$ at the point 1, defined within the instantaneous rest frame of the
particle 1 at $t'=$const [in constrast with the notation
$(F)_{\textstyle {}_1}$ used in (\ref{3}) for the old
regularization]. In addition, we shall introduce a
delta-pseudo-function denoted by ${\rm Pf}\Delta_1$ associated with
the new regularization [similarly to the delta-pseudo-function ${\rm
Pf}\delta_1$ which was defined in (\ref{14}) in the case of the old
regularization].

\subsection{Lorentz transformation of the source variables}

In this paper, it is sufficient to consider only those homogeneous
proper Lorentz transformations which change the velocity of a global
inertial frame $\{x^\mu\}=\{ct,{\bf x}\}$. More specifically, let us
consider the Lorentz boost

\begin{equation}\label{19}
x'^\mu=\Lambda^\mu_{~\nu}({\bf V})~\!x^\nu\;,
\end{equation}
where the Lorentz matrix $\Lambda^\mu_{~\nu}({\bf V})$, depending on
the constant boost velocity ${\bf V}$, is given by

\begin{mathletters}\label{20}\begin{eqnarray}
\Lambda^0_{~0}({\bf V})&=&\gamma\;,\\
\Lambda^i_{~0}({\bf V})&=&-\gamma {V^i\over c}\;,\\
\Lambda^0_{~j}({\bf V})&=&-\gamma {V_j\over c}\;,\\
\Lambda^i_{~j}({\bf V})&=&\delta^i_j+{\gamma^2\over 
\gamma+1}{V^i V_j\over c^2}\;.
\end{eqnarray}\end{mathletters}$\!\!$
We indifferently denote the components of the boost vector by ${\bf
V}=(V^i)=(V_i)$ (spatial indices $i,j=1,2,3$). The Lorentz factor
$\gamma$ reads

\begin{equation}\label{21}
\gamma={1\over \sqrt{1-{{\bf V}^2\over c^2}}}\;,
\end{equation}
with ${\bf V}^2=\delta_{ij}V^iV^j$ (of course $|{\bf V}|<c$). The
inverse transformation is $x^\nu=\Lambda^{~\nu}_{\mu}({\bf
V})~\!x'^\mu$ where the components of $\Lambda^{~\nu}_{\mu}({\bf
V})=\eta_{\mu\rho}\eta^{\nu\sigma}\Lambda^\rho_{~\sigma}({\bf V})$ are
obtained directly from (\ref{19}) by changing ${\bf V}\rightarrow
-{\bf V}$. The choice of sign made in the $0i$ components of the boost
(\ref{20}) is such that a particle which has velocity ${\bf V}$ at
time $t$ in the frame $\{x^\mu\}$ is at rest in the frame $\{x'^\mu\}$
at time $t'$.

We introduce on one side the space-time event ${\sc Q}$, which
represents for us a ``field'' point located outside the two world
lines of the particles, and on the other side the space-time events
${\sc P}_1$, ${\sc M}_1$ and ${\sc P}_2$, ${\sc M}_2$, which are
``source'' points, lying respectively on the world lines of the
particles 1 and 2 (see below for their definition).  The coordinates
of the event ${\sc Q}$ are $(t,{\bf x})$ in the frame $\{x^\mu\}$ and
$(t',{\bf x}')$ in the frame $\{x'^\mu\}$. Sorting out the spatial and
temporal indices in (\ref{19}), we have

\begin{mathletters}\label{22}\begin{eqnarray}
ct'&=&c\Lambda^0_{~0}t+\Lambda^0_{~j}x^j\;,\label{22a}\\
x'^i&=&c\Lambda^i_{~0}t+\Lambda^i_{~j}x^j\;.
\end{eqnarray}\end{mathletters}$\!\!$
The points ${\sc P}_1$ and ${\sc P}_2$ are now defined as the two
events that are located on the trajectories of the particles and are
``simultaneous'' with the event ${\sc Q}$ in the frame $\{x^\mu\}$,
i.e. that belong to the same spatial slice $t=$const as ${\sc Q}$. The
coordinates of ${\sc P}_1$ and ${\sc P}_2$ in $\{x^\mu\}$ are denoted
by $(t,{\bf y}_1)$ and $(t,{\bf y}_2)$ respectively, the two
trajectories ${\bf y}_1={\bf y}_1(t)$ and ${\bf y}_2={\bf y}_2(t)$
being parametrized by the coordinate time $t$ in that frame. On the
other hand, in the new frame $\{x'^\mu\}$, the coordinates of ${\sc
P}_1$ and ${\sc P}_2$ are $(\tau'_1,{\bf z}'_1)$ and $(\tau'_2,{\bf
z}'_2)$. Evidently, the primed coordinates are related to the unprimed
ones by the Lorentz boost (\ref{19}), so that

\begin{mathletters}\label{23}\begin{eqnarray}
c\tau'_1&=&c\Lambda^0_{~0}t+\Lambda^0_{~j}y^j_1\;,\\
{z'_1}^i&=&c\Lambda^i_{~0}t+\Lambda^i_{~j}y^j_1\;,
\end{eqnarray}\end{mathletters}$\!\!$
in the case of the event ${\sc P}_1$ [where $y^j_1 = y^j_1(t)$, 
$y^j_2 = y^j_2(t)$], and 

\begin{mathletters}\label{24}\begin{eqnarray}
c\tau'_2&=&c\Lambda^0_{~0}t+\Lambda^0_{~j}y^j_2\;,\\
{z'_2}^i&=&c\Lambda^i_{~0}t+\Lambda^i_{~j}y^j_2\;,
\end{eqnarray}\end{mathletters}$\!\!$
in the case of the event ${\sc P}_2$. In the new frame $\{x'^\mu\}$,
the source events that are simultaneous with ${\sc Q}$ are not ${\sc
P}_1$ and ${\sc P}_2$, but some other events ${\sc M}_1$ and ${\sc
M}_2$, whose coordinates in the primed frame are thus $(t',{\bf
y}'_1)$ and $(t',{\bf y}'_2)$; the coordinate time $t'$ is the same as
that of ${\sc Q}$ in the primed frame, and the spatial coordinates are
the trajectories of the particles ${\bf y}'_1={\bf y}'_1(t')$ and
${\bf y}'_2={\bf y}'_2(t')$ which are labelled by $t'$ in the new
frame. Let $(\tau_1,{\bf z}_1)$ and $(\tau_2,{\bf z}_2)$ be the
coordinates of ${\sc M}_1$ and ${\sc M}_2$ in the original frame
$\{x^\mu\}$. By definition,

\begin{mathletters}\label{25}\begin{eqnarray}
ct'&=&c\Lambda^0_{~0}\tau_1+\Lambda^0_{~j}z^j_1\;,\label{25a}\\
{y'_1}^i&=&c\Lambda^i_{~0}\tau_1+\Lambda^i_{~j}z^j_1\;,\\
ct'&=&c\Lambda^0_{~0}\tau_2+\Lambda^0_{~j}z^j_2\;,\\
{y'_2}^i&=&c\Lambda^i_{~0}\tau_2+\Lambda^i_{~j}z^j_2\;,
\end{eqnarray}\end{mathletters}$\!\!$
where ${y'_1}^i = {y'_1}^i(t')$ and ${y'_2}^i = {y'_2}^i(t')$. Beware
of our notation, where $\tau'_1$ (for instance) is the time coordinate
of ${\sc P}_1$ in $\{x'^\mu\}$ while $\tau_1$ is the time coordinate
in $\{x^\mu\}$ of the {\it different} event ${\sc M}_1$. Since the
events ${\sc M}_1$ and ${\sc M}_2$ are located on the world lines of
the particles parametrized by ${\bf y}_1(t)$ and ${\bf y}_2(t)$ in
$\{x^\mu\}$, it is clear that at time $\tau_1$ in that frame their
coordinates are related to the trajectories by

\begin{mathletters}\label{26}\begin{eqnarray}
{\bf z}_1&=&{\bf y}_1(\tau_1)\;,\label{26a}\\
{\bf z}_2&=&{\bf y}_2(\tau_2)\;.
\end{eqnarray}\end{mathletters}$\!\!$
Similarly, from the fact that ${\sc P}_1$ and ${\sc P}_2$ are also on
the world lines, which write as ${\bf y}'_1(t')$ and ${\bf y}'_2(t')$
in the frame $\{x'^\mu\}$, we deduce that their coordinates in
$\{x'^\mu\}$ satisfy

\begin{mathletters}\label{27}\begin{eqnarray}
{\bf z}'_1&=&{\bf y}'_1(\tau'_1)\;,\\
{\bf z}'_2&=&{\bf y}'_2(\tau'_2)\;.
\end{eqnarray}\end{mathletters}$\!\!$
By eliminating $t'$ from the equations (\ref{22a}) and (\ref{25a}) we
immediately obtain

\begin{equation}\label{28}
c\Lambda^0_{~0}(\tau_1-t)=\Lambda^0_{~i}(x^i-z_1^i)\;,
\end{equation}
or, equivalently, taking also into account (\ref{20}),

\begin{equation}\label{29}
\tau_1-t=-{1\over c^2}{\bf V}.({\bf x}-{\bf z}_1)\;,
\end{equation}
where the usual Euclidean scalar product between (boldface) vectors is
denoted by a dot. With the help of the latter formula for expressing
$\tau_1$, we can re-state the belonging of ${\bf z}_1$ to the particle
world line at time $\tau_1$ [see (\ref{26a})] as

\begin{equation}\label{30}
{\bf z}_1={\bf y}_1\!\!\left(t-{1\over c^2}{\bf V}.({\bf x}-{\bf z}_1)\right)\;.
\end{equation}
Recall that ${\bf z}_1$ is the spatial coordinate in the old frame of
the event ${\sc M}_1$ which is simultaneous with the field point ${\sc
Q}$ in the {\it new} frame. Clearly, the equation (\ref{30})
determines the vector ${\bf z}_1$ as a function of the coordinates
$(t,{\bf x})$ of the field-point event ${\sc Q}$ (see the
appendix). Here, let us view ${\bf z}_1$ as a ``vector'' field ${\bf
z}_1({\bf x})$, solution of (\ref{30}), lying in the three-dimensional
space $t=$const. It is evident from (\ref{30}) that the function ${\bf
z}_1({\bf x})$ admits a fixed point at ${\bf y}_1={\bf y}_1(t)$, in
the sense that

\begin{equation}\label{31}
{\bf z}_1({\bf y}_1)={\bf y}_1\;.
\end{equation}
Unless specified otherwise [like in (\ref{30})], the notation ${\bf
y}_1$ always means ${\bf y}_1(t)$. The mathematical justification of
(\ref{31}) is the following. From the fact that the world line of the
particle is time-like we can write, for any instants ${\hat t}$ and
${\bar t}$, the inequality $|{\bf y}_1({\hat t})-{\bf y}_1({\bar t})|<
c |{\hat t}-{\bar t}|$. Hence, applying the definition (\ref{30}), we
find that our function ${\bf z}_1({\bf x})$ obeys, for any positions
${\hat {\bf x}}$ and ${\bar {\bf x}}$, the further inequalities $|{\bf
z}_1({\hat {\bf x}})-{\bf z}_1({\bar {\bf x}})|< \case{1}{c}|{\bf
V}.({\hat {\bf x}}-{\bar {\bf x}})|\leq \case{|{\bf V}|}{c}|{\hat {\bf
x}}-{\bar {\bf x}}|$. Now recall that $\case{|{\bf V}|}{c}<1$, so the
latter inequalities mean exactly that the function ${\bf x}\rightarrow
{\bf z}_1({\bf x})$ is a {\it contracting} application with respect to
the usual Euclidean norm (i.e., it satisfies the property of Lipschitz
with a ratio $k=\case{|{\bf V}|}{c}$ strictly less than
one). Therefore, by the theorem of Picard, the function admits a
unique fixed point, which of course is nothing but ${\bf y}_1$.
(Besides, at the location of the fixed point, we have $\tau_1=t$.)

In this paper, we establish the general solution of the equation
(\ref{30}) in the form of an infinite (post-Newtonian) power series in
$1/c^2$. We shall not discuss the convergence properties of this
series and simply employ it to define the regularization up to any
relativistic order. This is sufficient for the application to the
problem of the equations of motion of particles in the post-Newtonian
approximation. The general solution of (\ref{30}), as determined in
the appendix, reads

\begin{equation}\label{32}
{\bf z}_1={\bf y}_1+\sum_{n=1}^{+\infty}{(-)^n\over c^{2n} n!}
\left(\partial\over \partial t\right)^{\!n-1}\!\!\Bigl[({\bf V}.{\bf r}_1)^n 
{\bf v}_1\Bigr]\;,
\end{equation}
with shorthand notations ${\bf y}_1={\bf y}_1(t)$, ${\bf r}_1={\bf
x}-{\bf y}_1(t)$ and ${\bf v}_1={\bf v}_1(t)$.  The many derivatives
$\partial/\partial t$ in the right side are partial time derivatives
with respect to the coordinate time $t$, the spatial coordinate ${\bf
x}$ being held constant. They act on ${\bf r}_1$ through the
trajectory ${\bf y}_1$: we have $\partial {\bf r}_1/\partial t=-{\bf
v}_1$ or $\partial ({\bf V}.{\bf r}_1)/\partial t=-{\bf V}.{\bf v}_1$
for instance. On the other side, they act of course on velocities and
(derivatives of) accelerations: thus $\partial {\bf v}_1/\partial
t={\bf a}_1$, $\partial {\bf a}_1/\partial t={\bf b}_1$, $\partial
{\bf b}_1/\partial t={\bf c}_1$, and so on, where ${\bf a}_1$, ${\bf
b}_1$, ${\bf c}_1$ represent the acceleration, and its first and
second derivatives (in these cases the partial derivative is a total
derivative, e.g. $d{\bf v}_1/dt={\bf a}_1$). Thus, to high
post-Newtonian order, (\ref{32}) contains many accelerations and
derivatives of accelerations, but it is understood that this formula
is order-reduced, consistently with the post-Newtonian order; i.e. all
accelerations and derivatives of accelerations are to be replaced by
the functionals of the positions and velocities deduced from the
equations of motion. Combining (\ref{29}) and (\ref{32}), we easily
find the corresponding solution for the time coordinate $\tau_1$,

\begin{equation}\label{33}
\tau_1=t+\sum_{n=1}^{+\infty}{(-)^n\over c^{2n} n!}\left(\partial
\over \partial t
\right)^{\!n-1}\!\!\Bigl[({\bf V}.{\bf r}_1)^n\Bigr]\;.
\end{equation}
[Of course, since ${\bf V}$ is a constant vector, it could be as well
put outside the partial time derivative operators in both (\ref{32})
and (\ref{33}).] Finally, equations (\ref{32}) and (\ref{33})
determine completely the space-time event ${\sc M}_1$. From them, we
can recover directly the fact that when ${\bf x}={\bf y}_1$ (at the
fixed point) then ${\bf z}_1={\bf y}_1$ and $\tau_1=t$: there are in
the right sides of both relations $n-1$ partial time derivatives
acting on a term that involves the $n$th power $({\bf V}.{\bf
r}_1)^n$, so that at least one of the scalar products ${\bf V}.{\bf
r}_1$ is left un-differentiated, and makes the sums in
(\ref{32})-(\ref{33}) vanish when ${\bf r}_1=0$.  Replacing both
${\bf z}_1$ and $\tau_1$ as given by the infinite post-Newtonian
series back into the relation (\ref{26a}), expressing both sides of
the resulting equation as the same type of post-Newtonian series with
the help of a formal Taylor expansion when $c\to\infty$, and finally
equating all the coefficients of these two series, yields an
interesting mathematical formula relating together some sums of
products of derivatives. This formula is derived in the appendix
(where we present also a direct proof of it). Notice that the same
reasoning as before can be done on the coordinates of the event ${\sc
P}_1$ in the new frame, that we find to be given by

\begin{mathletters}\label{33'}\begin{eqnarray}
{\bf z}'_1&=&{\bf y}'_1+\sum_{n=1}^{+\infty}{1\over c^{2n}
n!}\left(\partial\over \partial t'\right)^{\!n-1}\!\!\Bigl[({\bf
V}.{\bf r}'_1)^n {\bf v}'_1\Bigr]\;,\\
\tau'_1&=&t'+\sum_{n=1}^{+\infty}{1\over c^{2n} n!}\left(\partial\over \partial t'
\right)^{\!n-1}\!\!\Bigl[({\bf V}.{\bf r}'_1)^n\Bigr]\;,
\end{eqnarray}\end{mathletters}$\!\!$
where ${\bf y}'_1={\bf y}'_1(t')$, ${\bf r}'_1={\bf x}'-{\bf
y}'_1(t')$ and ${\bf v}'_1={\bf v}'_1(t')$. Evidently, the result
(\ref{33'}) can also be deduced directly from (\ref{32})-(\ref{33}) by
changing ${\bf V}$ into $-{\bf V}$ and replacing all the
non-primed variables by the corresponding primed ones.

We are now able to give all the transformation laws of field and
source variables between the frames $\{x^\mu\}$ and $\{x'^\mu\}$. Of
course, from (\ref{22}), the transformation of the field variables is
the standard Lorentz one,

\begin{mathletters}\label{34}\begin{eqnarray}
t'&=&\gamma \left(t-{1\over c^2}({\bf V}.{\bf x})\right)\;,\\ {\bf
x}'&=&{\bf x}-\gamma {\bf V} \left(t-{1\over c^2}{\gamma\over
\gamma+1}({\bf V}.{\bf x})\right)\;.\label{34b}
\end{eqnarray}\end{mathletters}$\!\!$
Concerning the source variables, we are interested in the expressions
of the new positions ${\bf y}'_1(t')$, ${\bf y}'_2(t')$ and velocities
${\bf v}'_1(t')$, ${\bf v}'_2(t')$ in the new frame at time
$t'$. These are straightforwardly obtained from inserting the results
(\ref{32}) and (\ref{33}) into the equations (\ref{25}), as well as
the similar results corresponding to the point 2. We find, for
trajectories,

\begin{mathletters}\label{35}\begin{eqnarray}
{\bf y}'_1={\bf y}_1&-&\gamma {\bf V} \left(t-{1\over c^2}{\gamma\over
\gamma+1}({\bf V}.{\bf x})\right)\nonumber\\
&+&\sum_{n=1}^{+\infty}{(-)^n\over c^{2n} n!}\left(\partial\over
\partial t\right)^{\!n-1}\!\!\left[({\bf V}.{\bf r}_1)^n \left({\bf
v}_1-{\gamma\over \gamma+1}{\bf V}\right)\right]\;,\\ {\bf y}'_2={\bf
y}_2&-&\gamma {\bf V} \left(t-{1\over c^2}{\gamma\over \gamma+1}({\bf
V}.{\bf x})\right)\nonumber\\ &+&\sum_{n=1}^{+\infty}{(-)^n\over
c^{2n} n!}\left(\partial\over \partial t\right)^{\!n-1}\!\!\left[({\bf
V}.{\bf r}_2)^n \left({\bf v}_2-{\gamma\over \gamma+1}{\bf
V}\right)\right]\;.
\end{eqnarray}\end{mathletters}$\!\!$
By subtracting the latter equations (\ref{35}) to ${\bf x}'$ as given
by (\ref{34b}) we obtain the spatial distances ${\bf r}'_1={\bf
x}'-{\bf y}'_1(t')$ and ${\bf r}'_2={\bf x}'-{\bf y}'_2(t')$ as

\begin{mathletters}\label{36}\begin{eqnarray}
{\bf r}'_1&=&{\bf r}_1-\sum_{n=1}^{+\infty}{(-)^n\over c^{2n}
n!}\left(\partial\over \partial t\right)^{\!n-1}\!\!\left[({\bf
V}.{\bf r}_1)^n \left({\bf v}_1-{\gamma\over \gamma+1}{\bf
V}\right)\right]\;,\\ {\bf r}'_2&=&{\bf
r}_2-\sum_{n=1}^{+\infty}{(-)^n\over c^{2n} n!}\left(\partial\over
\partial t\right)^{\!n-1}\!\!\left[({\bf V}.{\bf r}_2)^n \left({\bf
v}_2-{\gamma\over \gamma+1}{\bf V}\right)\right]\;.
\end{eqnarray}\end{mathletters}$\!\!$
These relations will play the crucial role in the definition of our
Lorentzian regularization. Of interest also is the expression of the
relative distance between the two particles, i.e. ${\bf y}'_{12}={\bf
y}'_1-{\bf y}'_2={\bf r}'_2-{\bf r}'_1$ given by

\begin{equation}\label{37}
{\bf y}'_{12}={\bf y}_{12}+\sum_{n=1}^{+\infty}{(-)^n\over c^{2n}
n!}\left(\partial\over \partial t\right)^{\!n-1}\!\!\left[({\bf
V}.{\bf r}_1)^n \left({\bf v}_1-{\gamma\over \gamma+1}{\bf
V}\right)-({\bf V}.{\bf r}_2)^n \left({\bf v}_2-{\gamma\over
\gamma+1}{\bf V}\right)\right]\;.
\end{equation}
Finally, we compute the expressions of the coordinate velocities ${\bf
v}'_1(t')=d{\bf y}'_1/dt'$ and ${\bf v}'_2(t')=d{\bf y}'_2/dt'$ in the
new frame. They follow immediately from the law of transformation of
the time derivative~: $\partial_t'=\gamma \partial_t+\gamma V^i
\partial_i$, and we obtain

\begin{mathletters}\label{38}\begin{eqnarray}
{\bf v}'_1&=&{1\over \gamma}{\bf v}_1-{\bf V}+{1\over
\gamma}\sum_{n=1}^{+\infty}{(-)^n\over c^{2n} n!}\left(\partial\over
\partial t\right)^{\!n}\!\!\left[({\bf V}.{\bf r}_1)^n \left({\bf
v}_1-{\gamma\over \gamma+1}{\bf V}\right)\right]\;,\label{38a}\\ {\bf
v}'_2&=&{1\over \gamma}{\bf v}_2-{\bf V}+{1\over
\gamma}\sum_{n=1}^{+\infty}{(-)^n\over c^{2n} n!}\left(\partial\over
\partial t\right)^{\!n}\!\!\left[({\bf V}.{\bf r}_2)^n \left({\bf
v}_2-{\gamma\over \gamma+1}{\bf V}\right)\right]\;.\label{38b}
\end{eqnarray}\end{mathletters}$\!\!$
Notice that although the velocities ${\bf v}'_1(t')$ and ${\bf
v}'_2(t')$ are some mere functions of the coordinate time $t'$ in the
new frame, they depend, when expressed in terms of quantities
belonging to the old frame, on both time {\it and} space coordinates
$t$ and ${\bf x}$. This is obvious because by changing the space
coordinate ${\bf x}$ of the field point ${\sc Q}$ while keeping
$t=$const we change the time coordinate $t'$ of the source events
${\sc M}_1$ and ${\sc M}_2$ and therefore the values of their particle
velocities (soon as the trajectories are accelerated). This fact is
important and has to be taken correctly into account in the
regularization process defined in the next subsection.  The inverse
formulas are obtained in the same way by substituting (\ref{33'}) into
the inverse of (\ref{23}). They correspond of course to changing ${\bf
V}$ into $-{\bf V}$ and replacing everywhere the un-primed labels by
primed ones. We find, for the spatial distances and velocities,

\begin{mathletters}\label{39}\begin{eqnarray}
{\bf r}_1&=&{\bf r}'_1-\sum_{n=1}^{+\infty}{1\over c^{2n}
n!}\left(\partial\over \partial t'\right)^{\!n-1}\!\!\left[({\bf
V}.{\bf r}'_1)^n \left({\bf v}'_1+{\gamma\over \gamma+1}{\bf
V}\right)\right]\;,\\ {\bf r}_2&=&{\bf
r}'_2-\sum_{n=1}^{+\infty}{1\over c^{2n} n!}\left(\partial\over
\partial t'\right)^{\!n-1}\!\!\left[({\bf V}.{\bf r}'_2)^n \left({\bf
v}'_2+{\gamma\over \gamma+1}{\bf V}\right)\right]\;,\\ {\bf
v}_1&=&{1\over \gamma}{\bf v}'_1+{\bf V}+{1\over
\gamma}\sum_{n=1}^{+\infty}{1\over c^{2n} n!}\left(\partial\over
\partial t'\right)^{\!n}\!\left[({\bf V}.{\bf r}'_1)^n \left({\bf
v}'_1+{\gamma\over \gamma+1}{\bf V}\right)\right]\;,\\ {\bf
v}_2&=&{1\over \gamma}{\bf v}'_2+{\bf V}+{1\over
\gamma}\sum_{n=1}^{+\infty}{1\over c^{2n} n!}\left(\partial\over
\partial t'\right)^{\!n}\!\left[({\bf V}.{\bf r}'_2)^n \left({\bf
v}'_2+{\gamma\over \gamma+1}{\bf V}\right)\right]\;.
\end{eqnarray}\end{mathletters}$\!\!$

\subsection{Definition of the regularization}

Let us consider a function $F$ belonging to the class ${\cal F}$ and
being at the same time a {\it scalar} under Lorentz transformations,
i.e. $F({\bf x},t)=F'({\bf x}',t')$. More precisely, we restrict
ourselves to the case of a dependence on ${\bf x}$ only via the
distances ${\bf r}_1$ and ${\bf r}_2$ [{\it cf} (\ref{18})]; this
means

\begin{equation}\label{48}
F[{\bf r}_1(t),{\bf r}_2(t);{\bf v}_1(t),{\bf v}_2(t)]=F'[{\bf
r}'_1(t'),{\bf r}'_2(t');{\bf v}'_1(t'),{\bf v}'_2(t');{\bf V}]\;,
\end{equation}
where we use the same slighly abusive notation as in (\ref{18}), with
addition, in the right side, of the explicit mention of the dependence
over the boost vector ${\bf V}$. All the variables in both frames
$\{x^\mu\}$ and $\{x'^\mu\}$ are related to each other by the formulas
developed in the previous subsection. The regularization process goes
as follows.

\bigskip\noindent
(I) Starting from $F[{\bf r}_1,{\bf r}_2;{\bf v}_1,{\bf v}_2]$ defined
in the frame $\{x^\mu\}$, we first determine the new functional
$F'[{\bf r}'_1,{\bf r}'_2;{\bf v}'_1,{\bf v}'_2;{\bf V}]$ in the
boosted frame $\{x'^\mu\}$. To do so, we replace all the variables
${\bf r}_1$, ${\bf r}_2$, ${\bf v}_1$, ${\bf v}_2$ by their
expressions in terms of the new ones ${\bf r}'_1$, ${\bf r}'_2$, ${\bf
v}'_1$, ${\bf v}'_2$ as given by the formulas (\ref{39}), in which it
is understood that all the accelerations are order-reduced up to some
given specified post-Newtonian order. Performing all the necessary
post-Newtonian re-expansions to that order, we indeed obtain in that
way (since $F$ is a Lorentz scalar) the new functional $F'$ of the new
distances ${\bf r}'_1$, ${\bf r}'_2$ and velocities ${\bf v}'_1$,
${\bf v}'_2$. In addition, $F'$ depends as expected on the constant
${\bf V}$ which is yet un-specified at this stage.

\bigskip\noindent
(II) We compute the Hadamard regularization of $F'$ at the point 1
following exactly the same rules as defined in (\ref{3}), but in the
boosted frame $\{x'^\mu\}$ (in particular, within the coordinate slice
$t'=$const). In words, we perform the expansion of $F'$ when the
spatial distance $r'_1$ tends to zero, and obtain the same type of
power-law expansion as in (\ref{1}) [since the form of the relations
(\ref{39}) shows that the structure of the expansions in both frames
must be the same]. However, we get some primed functional coefficients
${}_1f'_a$ that differ from the un-primed coefficients ${}_1f_a$
appearing in (\ref{1}). The boost vector ${\bf V}$ is simply held
constant in the process. Thus, $\forall N\in {\mathbb N}$,

\begin{equation}\label{49}
F'[{\bf r}'_1,{\bf r}'_2;{\bf v}'_1,{\bf v}'_2;{\bf V}]=\sum_{a_0\leq
a\leq N} {r'}_1^a \!\!\sous{1}{f'}_{a}\Bigl({\bf n}'_1;{\bf
y}'_{12};{\bf v}'_1,{\bf v}'_2;{\bf V}\Bigr)+o\left({r'}_1^N\right)\;,
\end{equation}
with the notation ${r'}_1=|{\bf x}'-{{\bf y}'}_1|$, ${\bf n}'_1=({\bf
x}'-{{\bf y}'}_1)/{r'}_1$ and ${\bf y}'_{12}={\bf y}'_1-{\bf
y}'_2$. (The fact that the coefficients ${}_1f'_a$ depend on ${\bf
y}'_{12}$ instead of the two individual trajectories ${\bf y}'_1$,
${\bf y}'_2$ is due to our restriction that $F'$ depends on ${\bf x}'$
via the distances ${\bf r}'_1$ and ${\bf r}'_2$; also, the
accelerations depend on the relative distance ${\bf y}'_{12}$.)  Now,
like in (\ref{3}), we pick up the zeroth-order coefficient in the
$r'_1$-expansion (\ref{49}) and average over the angles. This defines
a certain functional of the separation vector ${\bf y}'_{12}$, the
velocities ${\bf v}'_1$, ${\bf v}'_2$ and the boost velocity ${\bf
V}$,

\begin{equation}\label{50}
\sous{1}{\hat f'}_{0}\Bigl({\bf y}'_{12};{\bf v}'_1,{\bf v}'_2;{\bf V}\Bigr)
=\int {{d\Omega'}_1\over 4\pi} \!\!\sous{1}{f'}_{0}\Bigl({\bf
n'}_1;{\bf y}'_{12};{\bf v}'_1,{\bf v}'_2;{\bf V}\Bigr)\;.
\end{equation}
We insist that the angular average is performed {\it in the new
frame}, within the spatial hypersurface $t'=$const; in particular, the
solid angle element in (\ref{50}) is the one associated with the unit
direction ${\bf n}'_1$ in that hypersurface: ${d\Omega'}_1=d\Omega
({{\bf n}'}_1)$. Here again, ${\bf V}$ is considered as a simple
constant ``spectator'' vector during the average.

\bigskip\noindent
(III) We impose that the new frame is actually the rest frame of the
particle 1 at the event ${\sc P}_1$. Recalling that the Lorentz boost
(\ref{20}) brings a particle with velocity ${\bf V}$ in the frame
$\{x^\mu\}$ at rest in the frame $\{{x'}^\mu\}$, we see that we must
choose

\begin{equation}\label{51}
{\bf V}={\bf v}_1(t)\;.
\end{equation} 
We come back to the original variables in the un-primed frame by using
the transformation laws (\ref{37})-(\ref{38}), in the limit where the
field point ${\bf x}$ tends to the source point ${\bf y}_1(t)$
(because we are located at the event ${\sc P}_1$), with ${\bf V}={\bf
v}_1$ according to (\ref{51}). Note that, in this limit ${\bf r}_1\to
{\bf 0}$, the coordinate time $t'$ of the event ${\sc Q}$ in the primed
frame is equal to the coordinate time $\tau'_1$ of the event ${\sc
P}_1$. It is important to realize that both the computation of the
limit when ${\bf r}_1\to {\bf 0}$ and the replacement of the vector
${\bf V}$ by (\ref{51}) are to be done {\it after} performing the many
partial time differentiations in (\ref{37})-(\ref{38}).  Consider
first the primed variable ${\bf y}'_{12}$, which is given by
(\ref{37}) where we apply the replacement ${\bf r}_1={\bf 0}$ (as well
as ${\bf V}={\bf v}_1$). In (\ref{37}) the $n-1$ partial time
derivatives acting on the term proportional to $({\bf V}.{\bf r}_1)^n$
will clearly lead to zero in the limit ${\bf r}_1={\bf 0}$; indeed, by
an argument met previously, there are not ``enough'' derivatives to
make a non-zero contribution. So the variable to be used when coming
back to the original frame is

\begin{equation}\label{52}
{\bf y}'_{12}=\left({\bf y}_{12}-\sum_{n=1}^{+\infty}{(-)^n\over
c^{2n} n!}\left(\partial\over \partial t\right)^{\!n-1}\!\!\left[({\bf
V}.{\bf r}_2)^n \left({\bf v}_2-{\gamma\over \gamma+1}{\bf
V}\right)\right]\right)_{\!\!~\biggl|_{{\textstyle {{\bf r}_2={\bf
y}_{12}\atop \!\!{\bf V}={\bf v}_1}}\atop{}}}\;.
\end{equation}
As indicated by the notation one must implement the replacements of
${\bf r}_2$ by ${\bf y}_{12}$ (this is equivalent to ${\bf r}_1={\bf
0}$) and of ${\bf V}$ by ${\bf v}_1$ {\it after} the $n-1$ time
differentiations. In the case of the primed velocity of the particle
2, given by (\ref{38b}), we simply have

\begin{equation}\label{53}
{\bf v}'_2=\left({1\over \gamma}{\bf v}_2-{\bf V}+{1\over
\gamma}\sum_{n=1}^{+\infty}{(-)^n\over c^{2n} n!}\left(\partial\over
\partial t\right)^{\!n}\!\!\left[({\bf V}.{\bf r}_2)^n \left({\bf
v}_2-{\gamma\over \gamma+1}{\bf
V}\right)\right]\right)_{\!\!~\biggl|_{{\textstyle {{\bf r}_2={\bf
y}_{12}\atop \!\!{\bf V}={\bf v}_1}}\atop{}}}\;.
\end{equation}
The formulas (\ref{52}) and (\ref{53}) define, after order-reduction
of the accelerations, some functionals ${\bf y}'_{12}[{\bf
y}_{12};{\bf v}_1,{\bf v}_2]$ and ${\bf v}'_2[{\bf y}_{12};{\bf
v}_1,{\bf v}_2]$ that we use for coming back to the initial frame
$\{x^\mu\}$.  Clearly, the primed velocity ${\bf v}'_1$ of the point
1, at which we perform the regularization, deserves a special
treatment. From (\ref{38a}) we obtain

\begin{equation}\label{54}
{\bf v}'_1=\left({1\over \gamma}{\bf v}_1-{\bf V}+{1\over
\gamma}\sum_{n=1}^{+\infty}{(-)^n\over c^{2n} n!}\left(\partial\over
\partial t\right)^{\!n}\!\!\left[({\bf V}.{\bf r}_1)^n \left({\bf
v}_1-{\gamma\over \gamma+1}{\bf
V}\right)\right]\right)_{\!\!~\biggl|_{{\textstyle {\!{\bf r}_1={\bf
0}\atop {\bf V}={\bf v}_1}}\atop{}}}\;.
\end{equation}
Here, there are $n$ time derivatives which is {\it a priori} enough to
make a contribution. The only possibility is to differentiate
successively each of the $n$ factors ${\bf V}.{\bf r}_1$, yielding for
each of the terms in the sum $n!$ identical contributions. Hence, we
arrive at a much simpler series,

\begin{equation}\label{54'}
{\bf v}'_1=\left({1\over \gamma}{\bf v}_1-{\bf V}+{1\over
\gamma}\sum_{n=1}^{+\infty}\left({{\bf V}.{\bf v}_1\over
c^{2}}\right)^n\!\!\left({\bf v}_1-{\gamma\over \gamma+1}{\bf
V}\right)\right)_{{\textstyle \!\!~\biggl|_{{\bf V}={\bf v}_1}}}\;,
\end{equation}
which can now easily be summed up. The result is

\begin{equation}\label{54''}
{\bf v}'_1=\left({{1\over \gamma}{\bf v}_1-{\bf V}+{\gamma\over
\gamma+1}{{\bf V}.{\bf v}_1\over c^{2}}{\bf V}\over 1-{{\bf V}.{\bf
v}_1\over c^{2}}}
\right)_{{\textstyle \!\!~\biggl|_{{\bf V}={\bf v}_1}}}\;,
\end{equation}
from which we immediately deduce that the primed velocity of the
particle 1 must be zero,

\begin{equation}\label{55}
{\bf v}'_1={\bf 0}\;.
\end{equation} 
This is of course the expected result because the boost velocity was
chosen to be equal to the instantaneous velocity of the particle 1 in
the un-primed frame at the instant $t$; however, the details of the
above proof constitute a necessary consistency check of the formulas.

\bigskip\noindent 
(IV) The choice of boost vector ${\bf V}={\bf v}_1$,
together with the equivalent statement that ${\bf v}'_1={\bf 0}$, as
well as the expressions (\ref{52}) and (\ref{53}) defining the two
functionals ${\bf y}'_{12}[{\bf y}_{12};{\bf v}_1,{\bf v}_2]$ and
${\bf v}'_2[{\bf y}_{12};{\bf v}_1,{\bf v}_2]$, are put into
(\ref{50}), which gave the result ${}_1{\hat f'}_{0}$ of the spherical
average in the Hadamard regularization performed in the primed
frame. Therefore, the regularized value of $F$ at the point 1 is
defined by

\begin{equation}\label{56}
[F]_{\textstyle {}_1}=\sous{1}{\hat f'}_{0}\Bigl({\bf y}'_{12}[{\bf
y}_{12};{\bf v}_1,{\bf v}_2];{\bf 0},{\bf v}'_2[{\bf y}_{12};{\bf
v}_1,{\bf v}_2];{\bf v}_1\Bigr)\;.
\end{equation}
The new regularization $[F]_{\textstyle {}_1}$ acts, like the old one
$(F)_{\textstyle {}_1}$, as a certain functional of the relative
distance ${\bf y}_{12}$ and the velocities ${\bf v}_1$, ${\bf
v}_2$. However, in generic cases, $[F]_{\textstyle {}_1}$ differs from
$(F)_{\textstyle {}_1}$ by relativistic terms at least of the order
$1/c^2$ [we investigate in Section \ref{IV} the exact relation between
both regularizations to the first relativistic order $1/c^2$]. In the
problem of the post-Newtonian equations of motion, we have found
\cite{BF00,BFeom} that the new regularization $[F]_{\textstyle {}_1}$
adds some extra terms to the acceleration computed using the 
regularization $(F)_{\textstyle {}_1}$; these new terms are of 
order 3PN and manage to make the 3PN equations of motion invariant with
respect to Lorentz transformations. Indeed, with the regularization
$(F)_{\textstyle {}_1}$ the Lorentz invariance of the equations of
motion would be broken at the 3PN order. Finally, let us introduce as
we did in \cite{BFreg} (see also Section \ref{II}) 
a delta-pseudo-function associated with the new
regularization $[F]_{\textstyle {}_1}$. By definition, the
``Lorentzian'' delta-pseudo-function denoted ${\rm Pf}\Delta_1$
[to contrast with the non-invariant one ${\rm Pf}\delta_1$ defined by
(\ref{14})] is such that
 
\begin{equation}\label{57}
\forall F\in {\cal F},\quad
<{\rm Pf}\Delta_1,F>=[F]_{\textstyle {}_1} \;,
\end{equation}
where the right side is given by the new regularization (\ref{56}). By
definition, we have in the case of the new regularization the same
laws for the multiplication as in Section \ref{II}, for instance

\begin{equation}\label{58}
FG~\!.~\!{\rm Pf}\Delta_1={\rm Pf}(F\Delta_1)~\!.~\!G={\rm
Pf}(F\Delta_1)~\!.~\!{\rm Pf}G={\rm Pf}(FG\Delta_1)\;,
\end{equation}
where the pseudo-function ${\rm Pf}(F\Delta_1)$ is defined by 

\begin{equation}\label{59}
\forall G\in {\cal F}\;,\quad
<{\rm Pf}(F\Delta_1),G>=[FG]_{\textstyle {}_1}\;.
\end{equation}
This pseudo-function ${\rm Pf}(F\Delta_1)$ is at the basis of our
proposal for the stress-energy tensor of point-particles in Section
\ref{V}. And, like in the case of ${\rm Pf}(F\delta_1)$, we are not
allowed to replace this pseudo-function by the product of the
regularized value of the function times the delta-pseudo-function,
namely

\begin{equation}\label{60}
{\rm Pf}(F\Delta_1)\not= [F]_{\textstyle {}_1}~{\rm Pf}\Delta_1\;.
\end{equation}

The derivatives of ${\rm Pf}\Delta_1$ and ${\rm Pf}(F\Delta_1)$ are
constructed in the same way as for the original regularization in
Section \ref{II}. Therefore,

\begin{equation}\label{60a}
\forall G\in{\cal F}\;,\qquad <\partial_i[{\rm Pf}(F\Delta_1)],G>
=-<{\rm Pf}(F\Delta_1),\partial_iG>=-[F\partial_iG]_{\textstyle {}_1}\;.
\end{equation}
However, the identity (\ref{5}) is not valid in the case of the new
regularization, so we do not have a result similar to (\ref{16c}) [see
(\ref{75}) for the equivalent of (\ref{5}) at the first relativistic
order]. For the product of ${\rm Pf}(F\delta_1)$ with some ${\rm
Pf}G$, the Leibniz rule holds:

\begin{equation}\label{60c}
\partial_i[{\rm Pf}(F\Delta_1)~\!\!.~\!{\rm Pf}G]=
\partial_i[{\rm Pf}(F\Delta_1)]~\!\!.~\!{\rm Pf}G+{\rm Pf}
(F\Delta_1)~\!\!.~\!\partial_i({\rm Pf}G)\;.
\end{equation}
This is a consequence of the definition (\ref{60a}) and the law
(\ref{58}).
 
\section{The regularization at the first relativistic order}\label{IV}

At this point, it is instructive (and useful in practice) to present
the complete formulas that define the Lorentzian regularization
$[F]_{\textstyle {}_1}$ at the level of the first relativistic
corrections $1/c^2$, i.e. neglecting all the terms of order
$O(1/c^4)$. [Notice that, consistently with Section \ref{III}, we must
consider that the boost vector ${\bf V}$ itself is of order $O(1)$, so
that for instance the factor ${\bf V}^2/c^2$ really represents a small
relativistic correction of the order $O(1/c^2)$.] Furthermore, we
shall obtain at this $1/c^2$ level a formula linking the new
regularization $[F]_{\textstyle {}_1}$ to the old one $(F)_{\textstyle
{}_1}$. Like in Section \ref{III}, we assume that the function $F$
depends on ${\bf x}$ through the two distances ${\bf r}_1(t)$ and
${\bf r}_2(t)$ only; this implies a relation between the partial
derivatives:

\begin{equation}\label{61}
\partial_iF+{\partial F\over \partial y_1^i}+{\partial F\over \partial y_2^i}=0
\end{equation}
(where $\partial_i=\partial/\partial x^i$). We suppose also
that $F$ is a Lorentz scalar, {\it cf} (\ref{48}).

We follow the general specification for the regularization in Section
\ref{III}. We first express the vectorial distances ${\bf r}_1$, ${\bf
r}_2$ and velocities ${\bf v}_1$, ${\bf v}_2$ in the boosted frame
$\{x'^\mu\}$ using the transformation formulas (\ref{39}) restricted
to the order $1/c^2$. For the distances, we get

\begin{mathletters}\label{62}\begin{eqnarray}
{\bf r}_1&=&{\bf r}'_1-{1\over c^2}({\bf V}.{\bf r}'_1)\Bigl[{\bf
v}'_1+\case{1}{2}{\bf V}\Bigr]+O\left({1\over c^4}\right)\;,\\ {\bf
r}_2&=&{\bf r}'_2-{1\over c^2}({\bf V}.{\bf r}'_2)\Bigl[{\bf
v}'_2+\case{1}{2}{\bf V}\Bigr]+O\left({1\over c^4}\right)\;.
\end{eqnarray}\end{mathletters}$\!\!$
The relative distance ${\bf y}_{12}={\bf r}_2-{\bf r}_1$ reads as
 
\begin{equation}\label{63}
{\bf y}_{12}={\bf y}'_{12}+{1\over c^2}\Bigl[-\case{1}{2}({\bf V}.{\bf
y}'_{12}){\bf V}+({\bf V}.{\bf r}'_1){\bf v}'_1-({\bf V}.{\bf
r}'_2){\bf v}'_2\Bigr]+O\left({1\over c^4}\right)\;,
\end{equation}
while, for instance, the relative separation $r_{12}=|{\bf y}_{12}|$
is

\begin{equation}\label{64}
r_{12}=r'_{12}\left(1+{1\over c^2}\Bigl[-\case{1}{2}({\bf V}.{\bf
n}'_{12})^2+\case{r'_1}{r'_{12}}({\bf V}.{\bf n}'_1)({\bf v}'_1.{\bf
n}'_{12})-\case{r'_2}{r'_{12}}({\bf V}.{\bf n}'_2)({\bf v}'_2.{\bf
n}'_{12})\Bigr]\right) +O\left({1\over c^4}\right)\;,
\end{equation}
where ${\bf n}'_{1}={\bf r}'_{1}/r'_{1}$, ${\bf n}'_{2}={\bf
r}'_{2}/r'_{2}$ and ${\bf n}'_{12}={\bf y}'_{12}/r'_{12}$. For the two
velocities, we find

\begin{mathletters}\label{65}\begin{eqnarray}
{\bf v}_1&=&{\bf v}'_1+{\bf V}+{1\over
c^2}\left(\Bigl[-\case{1}{2}{\bf V}^2-{\bf V}.{\bf v}'_1\Bigl]{\bf
v}'_1-\case{1}{2}({\bf V}.{\bf v}'_1){\bf V}+({\bf V}.{\bf r}'_1) {\bf
a}'_1\right)+O\left({1\over c^4}\right)\;,\\ {\bf v}_2&=&{\bf v}'_2+{\bf
V}+{1\over c^2}\left(\Bigl[-\case{1}{2}{\bf V}^2-{\bf V}.{\bf
v}'_2\Bigl]{\bf v}'_2-\case{1}{2}({\bf V}.{\bf v}'_2){\bf V}+({\bf
V}.{\bf r}'_2) {\bf a}'_2\right)+O\left({1\over c^4}\right)\;,
\end{eqnarray}\end{mathletters}$\!\!$
where the two accelerations ${\bf a}'_1$ and ${\bf a}'_2$ are to be
replaced, consistently with the approximation, by their Newtonian
values~: ${\bf a}'_1=-{G m_2\over {r'}_{12}^2}{\bf
n}'_{12}+O\left({1\over c^2}\right)$ and ${\bf a}'_2={G m_1\over
{r'}_{12}^2}{\bf n}'_{12}+O\left({1\over c^2}\right)$. [Notice that in
Section \ref{III} the regularization has been defined regardless of
the type of special-relativistic interaction involved; in the case of
electromagnetism, for instance, we should simply replace the
accelerations by their Coulombian values in (\ref{65}).]

Next, we substitute the expressions (\ref{62}) and (\ref{65}) into the
scalar function $F[{\bf r}_1,{\bf r}_2;{\bf v}_1,{\bf v}_2]$ and
perform the expansion to the first order. The result is the scalar
function $F'[{\bf r}'_1,{\bf r}'_2;{\bf v}'_1,{\bf v}'_2;{\bf V}]$ in
the new frame; thus

\begin{eqnarray}\label{67}
&&F'[{\bf r}'_1,{\bf r}'_2;{\bf v}'_1,{\bf v}'_2;{\bf V}]=F[{\bf
r}'_1,{\bf r}'_2;{\bf v}'_1+{\bf V},{\bf v}'_2+{\bf V}]\nonumber\\
&&\qquad\qquad+{1\over c^2}({\bf V}.{\bf
r}'_1)\Bigl[{v'}^i_1+\case{1}{2}V^i\Bigr]{\partial F\over \partial
y_1^i} +{1\over c^2}({\bf V}.{\bf
r}'_2)\Bigl[{v'}^i_2+\case{1}{2}V^i\Bigr]{\partial F\over \partial
y_2^i}\nonumber\\ &&\qquad\qquad+{1\over
c^2}\left(\Bigl[-\case{1}{2}{\bf V}^2-{\bf V}.{\bf
v}'_1\Bigl]{v'}^i_1-\case{1}{2}({\bf V}.{\bf v}'_1)V^i+({\bf V}.{\bf
r}'_1) {a'}^i_1\right){\partial F\over \partial v_1^i}\nonumber\\
&&\qquad\qquad+{1\over c^2}\left(\Bigl[-\case{1}{2}{\bf V}^2-{\bf
V}.{\bf v}'_2\Bigl]{v'}^i_2-\case{1}{2}({\bf V}.{\bf v}'_2)V^i+({\bf
V}.{\bf r}'_2) {a'}^i_2\right){\partial F\over \partial
v_2^i}+O\left({1\over c^4}\right)\;,
\end{eqnarray}
where we have used $\case{\partial F}{\partial r_1^i}=-\case{\partial
F}{\partial y_1^i}$ and $\case{\partial F}{\partial
r_2^i}=-\case{\partial F}{\partial y_2^i}$. Note that, to this order,
the partial derivatives in (\ref{67}) can be evaluated at the primed
values ${\bf r}'_1$, ${\bf r}'_2$ and ${\bf v}'_1+{\bf V}$, ${\bf
v}'_2+{\bf V}$, or equivalently at the non-primed ones ${\bf r}_1$,
${\bf r}_2$ and ${\bf v}_1$, ${\bf v}_2$. Now we pick up in the new
frame the term of zeroth order in the expansion when $r'_1\to 0$, and
perform the angular average with respect to the direction ${\bf
n}'_1$. This yields the functional of the variables ${\bf y}'_{12}$,
${\bf v}'_1$, ${\bf v}'_2$ and ${\bf V}$ which has been defined in
(\ref{50}). Since these operations of expanding and averaging
represent nothing but the Hadamard regularization in the old sense of
(\ref{3}), we can denote them by using the parenthesis appropriate for
this regularization. Therefore,

\begin{eqnarray}\label{68}
&&\sous{1}{\hat f'}_{0}\Bigl({\bf y}'_{12};{\bf v}'_1,{\bf v}'_2;{\bf
V}\Bigr) =\Biggl(~F[{\bf r}_1,{\bf r}_1+{\bf y}'_{12};{\bf v}'_1+{\bf
V},{\bf v}'_2+{\bf V}]\nonumber\\ &&\quad\quad+{1\over c^2}({\bf
V}.{\bf r}_1)\Bigl[{v'}^i_1+\case{1}{2}V^i\Bigr]{\partial F\over
\partial y_1^i} +{1\over c^2}\Bigl({\bf V}.{\bf r}_1+{\bf V}.{\bf
y}'_{12}\Bigr)\Bigl[{v'}^i_2+\case{1}{2}V^i\Bigr]{\partial F\over
\partial y_2^i}\nonumber\\ &&\quad\quad+{1\over
c^2}\left(\Bigl[-\case{1}{2}{\bf V}^2-{\bf V}.{\bf
v}'_1\Bigl]{v'}^i_1-\case{1}{2}({\bf V}.{\bf v}'_1)V^i+({\bf V}.{\bf
r}_1) {a'}^i_1\right){\partial F\over \partial v_1^i}\nonumber\\
&&\quad\quad+{1\over c^2}\left(\Bigl[-\case{1}{2}{\bf V}^2-{\bf
V}.{\bf v}'_2\Bigl]{v'}^i_2-\case{1}{2}({\bf V}.{\bf
v}'_2)V^i+\Bigl[{\bf V}.{\bf r}_1+{\bf V}.{\bf y}'_{12}\Bigr]
{a'}^i_2\right){\partial F\over \partial
v_2^i}~\Biggr)_1+O\left({1\over c^4}\right)\;.
\end{eqnarray}
We have replaced here the vectorial distance ${\bf r}'_1$ by the
un-primed notation ${\bf r}_1$, noticing that ${\bf r}'_1$ is the
dummy variable with respect to which the regularization proceeds (with
this notation ${\bf r}'_2$ is replaced by ${\bf r}_1+{\bf
y}'_{12}$). Following (\ref{56}), the Lorentzian regularization
$[F]_{\textstyle {}_1}$ is achieved by posing ${\bf V}={\bf v}_1$ and
${\bf v}'_1={\bf 0}$, as well as ${\bf y}'_{12}={\bf y}'_{12}[{\bf
y}_{12};{\bf v}_1,{\bf v}_2]$ and ${\bf v}'_2={\bf v}'_2[{\bf
y}_{12};{\bf v}_1,{\bf v}_2]$, where the latter functionals are
defined in the general case by (\ref{52}) and (\ref{53}). It is
convenient to obtain first an intermediate formula by setting ${\bf
V}={\bf v}_1$ and ${\bf v}'_1={\bf 0}$ in (\ref{68}), and by replacing
into the terms that are already of order $1/c^2$ the primed variables
${\bf y}'_{12}$ and ${\bf a}'_1$, ${\bf a}'_2$ by the un-primed
ones. Using also the identity (\ref{61}), we arrive at

\begin{eqnarray}\label{69}
[F]_{\textstyle {}_1} =\Biggl(&F&[{\bf r}_1,{\bf r}_1+{\bf
y}'_{12};{\bf v}_1,{\bf v}_1+{\bf v}'_2] +{1\over 2c^2}({\bf v}_1.{\bf
r}_1) v^i_1\partial_i F\nonumber\\ &+&{1\over c^2}({\bf v}_1.{\bf
r}_1)\biggl[v_1^i{\partial F\over \partial y_1^i} +v_2^i{\partial
F\over \partial y_2^i} +a^i_1{\partial F\over \partial
v_1^i}+a^i_2{\partial F\over \partial v_2^i}\biggr]\nonumber\\
&+&{1\over c^2}\left(\case{1}{2}({\bf v}_1.{\bf v}_2)v^i_1
+\Bigl[\case{1}{2}{\bf v}_1^2-{\bf v}_1.{\bf v}_2\Bigl]v^i_2 +({\bf
v}_1.{\bf y}_{12})a^i_2\right){\partial F\over \partial
v_2^i}\nonumber\\ &+&{1\over c^2}({\bf v}_1.{\bf
y}_{12})\Bigl[-\case{1}{2}v^i_1+v_2^i\Bigr]{\partial F\over \partial
y_2^i}~\Biggr)_1 +O\left({1\over c^4}\right)\;,
\end{eqnarray}
where ${\bf y}'_{12}$ and ${\bf v}'_2$ in the first term of the right
side are given functions of ${\bf y}_{12}$, ${\bf v}_1$ and ${\bf
v}_2$ obtained by approximating (\ref{52}) and (\ref{53}) to the first
order; we find

\begin{mathletters}\label{70}\begin{eqnarray}
{\bf y}'_{12}&=&{\bf y}_{12}+{1\over c^2}({\bf v}_1.{\bf
y}_{12})\Bigl[-\case{1}{2}{\bf v}_1+{\bf v}_2\Bigr]+O\left({1\over
c^4}\right)\;,\\ {\bf v}'_2&=&-{\bf v}_1+{\bf v}_2+{1\over
c^2}\left(-\case{1}{2}({\bf v}_1.{\bf v}_2){\bf v}_1
+\Bigl[-\case{1}{2}{\bf v}_1^2+{\bf v}_1.{\bf v}_2\Bigl]{\bf
v}_2-({\bf v}_1.{\bf y}_{12}) {\bf a}_2\right)+O\left({1\over
c^4}\right)
\end{eqnarray}\end{mathletters}$\!\!$
(where the acceleration is equal to its Newtonian value). By inserting
(\ref{70}) into (\ref{69}) and expanding to order $1/c^2$, it is
easily seen that we cancel out exactly the two last terms in the
right-hand side of (\ref{69}), so that the result simplifies
appreciably:

\begin{eqnarray}\label{71}
[F]_{\textstyle {}_1} =\Biggl(&F&[{\bf r}_1,{\bf r}_2;{\bf v}_1,{\bf
v}_2] +{1\over 2c^2}({\bf v}_1.{\bf r}_1) v^i_1\partial_i F\nonumber\\
&+&{1\over c^2}({\bf v}_1.{\bf r}_1)\biggl[v_1^i{\partial F\over
\partial y_1^i} +v_2^i{\partial F\over \partial y_2^i} +a^i_1{\partial
F\over \partial v_1^i}+a^i_2{\partial F\over \partial v_2^i}\biggr]
~\Biggr)_1 +O\left({1\over c^4}\right)\;.
\end{eqnarray}
Finally, we recognize in the right side the partial time-derivative,

\begin{equation}\label{73}
\partial_tF=v_1^i{\partial F\over \partial y_1^i}
+v_2^i{\partial F\over \partial y_2^i} +a^i_1{\partial F\over \partial
v_1^i}+a^i_2{\partial F\over \partial v_2^i}\;,
\end{equation}
so that our final result writes

\begin{equation}\label{74}
[F]_{\textstyle {}_1}=\Biggl(F+{1\over c^2}({\bf r}_1.{\bf
v}_1)\Bigl[\partial_tF+\case{1}{2}v^i_1\partial_i
F\Bigr]\Biggr)_1+O\left({1\over c^4}\right)\;.
\end{equation}

The result (\ref{74}) displays the first relativistic corrections
brought about by our Lorentzian regularization $[F]_{\textstyle
{}_1}$. As a check of the formula, let us apply it to the case of the
special combination $\partial_iF-3\case{n_1^i}{r_1}F$ which, as we
know from (\ref{5}), has no partie finie at the point 1 in the sense
of the old regularization. This is no longer true in the sense of the
new regularization. Using the equation (\ref{74}) we find instead

\begin{equation}\label{75}
[\partial_iF]_{\textstyle {}_1}=\Biggl[3{n_1^i\over
r_1}\biggl(1-{1\over c^2}({\bf n}_1.{\bf v}_1)^2\biggr)F-{1\over
c^2}v_1^i\partial_tF\Biggr]_1+O\left({1\over c^4}\right)\;.
\end{equation}
The check consists of remarking that because of (\ref{5}) we have
$(\partial'_iF'-3{{n'}_1^i\over r'_1}F')_{\textstyle {}_1}=0$ in the
rest frame of the particle 1, therefore the equation
$[\partial'_iF'-3{{n'}_1^i\over r'_1}F']_{\textstyle {}_1}=0$ must
hold in any frame by definition of the new regularization. In the
frame where the particle velocity is ${\bf v}_1$ we have ${\bf
r}'_1={\bf r}_1+\case{1}{2c^2}({\bf v}_1.{\bf r}_1){\bf
v}_1+O(\case{1}{c^4})$ and
$\partial'_i=\partial_i+\case{1}{c^2}v_1^i\partial_t+
\case{1}{2c^2}v_1^iv_1^j\partial_j
+O(\case{1}{c^4})$. Inserting these relations into the previous
equation, and using the fact that $F$ is a scalar, we recover the
formula (\ref{75}) after a short computation.

\section{The stress-energy tensor of point-particles}\label{V}

With the Lorentzian regularization in hands, we make a proposal for
the description of point-like particles in (post-Newtonian
approximations of) general relativity. We recall first the general
context of the problem. We want to solve the field equations of
general relativity by means of analytic post-Newtonian series, with
matter source describing appropriately defined point-particles. The
stress-energy tensor of the matter source is supposed to be spatially
isolated; we recall that, in this case, general relativity admits the
Poincar\'e group as a global symmetry. We assume the existence and
unicity of a global harmonic coordinate system, defined by the gauge
conditions

\begin{mathletters}\label{77}\begin{eqnarray}
&&\partial_\nu h^{\mu\nu} = 0\;,\\
&&h^{\mu\nu} = \sqrt{g} g^{\mu\nu}-
\eta^{\mu\nu}\;,
\end{eqnarray}\end{mathletters}$\!\!$
where $g^{\mu\nu}$ denotes the inverse of the covariant metric
$g_{\mu\nu}$, and where $g$ is the opposite of its determinant. The
harmonic gauge conditions (\ref{77}) introduce a preferred Minkowskian
structure, with Minkowski metric given by $\eta^{\mu\nu}={\rm
diag}(-1,1,1,1)=\eta_{\mu\nu}$. Thus, the gravitational field can be
described in harmonic coordinates by the Lorentzian tensor field
$h^{\mu\nu}$ propagating on the Minkowskian background
$\eta^{\mu\nu}$. Similarly, one can think of the trajectories of the
particles as accelerated world lines in Minkowski space-time. Subject
to the conditions (\ref{77}) the Einstein field equations take the
form of wave equations on the flat background,

\begin{equation}\label{78}
\Box h^{\mu\nu} = \case{16\pi G}{c^4} g~\!T^{\mu\nu} + 
\Lambda^{\mu\nu}[h,\partial h,\partial^2h]\;,
\end{equation} 
where the flat d'Alembertian operator is given by
$\Box=\eta^{\mu\nu}\partial_\mu\partial_\nu$. The right-hand side is
made of the sum of the matter source term, with spatially compact
support, plus the gravitational source term $\Lambda^{\mu\nu}$, given
by a certain functional of the field variables $h^{\rho\sigma}$ and
its first and second space-time derivatives, and at least of second
order in $h$. A consequence of the harmonicity conditions is that

\begin{equation}\label{78'}
\partial_\nu\left(g~\!T^{\mu\nu} +\case{c^4}{16\pi G} 
\Lambda^{\mu\nu}\right)=0\;,
\end{equation} 
which is equivalent (through the contracted Bianchi identity) to the
covariant conservation of the matter stress-energy tensor
$T^{\mu\nu}$,

\begin{equation}\label{79}
\nabla_\nu T^{\mu\nu}=0\;,
\end{equation}
the latter equation being in turn equivalent to 

\begin{equation}\label{79'}
\partial_\nu\left(\sqrt{g}~\! g_{\lambda\mu}~\!T^{\mu\nu}\right)
=\case{1}{2}\sqrt{g}~\!\partial_\lambda g_{\mu\nu}~\!T^{\mu\nu}\;.
\end{equation}
In this section we regard the matter tensor $T^{\mu\nu}$ as a Lorentz
tensor defined with respect to the Minkowski metric $\eta_{\mu\nu}$
singled out by our choice of harmonic coordinates.

To define a model for point-like particles, we follow essentially the
derivation of the stress-energy tensor of {\it test} masses moving on
a fixed {\it smooth} background (see e.g. Weinberg \cite{Weinberg}
p. 360). However, in the case of ``self-gravitating'' particles, we do
not have a smooth background at our disposal, and the metric becomes
singular at the location of the point-masses. Essentially, we shall
propose the value of the (post-Newtonian) metric coefficients on each
of the particles to be given by the Lorentzian regularization defined
in Section
\ref{III}. This entails supposing that the metric coefficients belong
to the class of functions ${\cal F}$. This is correct up to the 2PN
order \cite{BFP98}; however, we know that the expansion of the metric
coefficients (in harmonic coordinates) near the particles, instead of
being of the type (\ref{1})-(\ref{2}), involve some logarithms of the
distance to the singularities starting at 3PN order. It was shown
\cite{BF00} that, at this order, the logarithms can be considered
as some constants and included into the definition of the partie
finie; moreover, they can be finally eliminated from the equations of
motion by a change of coordinates. This suggests that we might
consider more generally the logarithms as some constants, motivating
our assumption that $g_{\mu\nu}\in {\cal F}$. On the other hand, it is
known \cite{BF00,BFeom} that the constants $s_1$ and $s_2$ entering the
partie-finie integral (\ref{6}) must be adjusted in order that the
equations of motion can be deducible from a Lagrangian, and in
particular admit a conserved energy. For these reasons (presence
of logarithms, equations of motion not directly admitting an energy),
the following derivation of the stress-energy tensor for particles
cannot be considered to be a rigorous proof. However, as we shall see,
it is nicely consistent with the regularization, and its result
satisfying. Our basic assumption is that the dynamics of the particles
follows from the variation, with respect to the metric, of the action

\begin{equation}\label{80}
I_{\rm particle} = -m_1 c \int_{-\infty}^{+\infty}
dt~\!\sqrt{-[g_{\mu\nu}]_{\textstyle {}_1} v_1^\mu v_1^\nu} +
1\leftrightarrow 2 \;,
\end{equation}
where $v_1^\mu = (c,d {\bf y}_1 /dt)$ denotes the coordinate velocity
of the particle 1 (we consider a two-body system, but the
generalization to $N$ bodies is immediate). The crucial point is that
the value of $g_{\mu\nu}$ at 1 is assumed to be given by the
Lorentzian regularization defined in Section \ref{III}. We vary the
action (\ref{80}) with respect to the metric, i.e. we imagine that
$g_{\mu\nu}\in {\cal F}$ is subject to an infinitesimal variation
$g_{\mu\nu}\rightarrow g_{\mu\nu}+\delta g_{\mu\nu}$ and compute the
corresponding change in the action. However, we want the variation of
the metric to correspond to the same matter system with two
singularities 1 and 2. The evident and most natural way to ensure this
is to suppose that $\delta g_{\mu\nu}\in {\cal F}$. Under the latter
variation the regularized value of the metric at the point 1 undergoes
the infinitesimal change $[g_{\mu\nu}]_{\textstyle {}_1}\rightarrow
[g_{\mu\nu}]_{\textstyle {}_1}+[\delta g_{\mu\nu}]_{\textstyle
{}_1}$. Therefore, the variation of the action (\ref{80}) reads as

\begin{equation}\label{81}
\delta I_{\rm particle} = \case{1}{2} m_1 c \int_{-\infty}^{+\infty} 
dt~\!{ v_1^\mu v_1^\nu \over \sqrt{-[g_{\rho\sigma}]_{\textstyle {}_1}
v_1^\rho v_1^\sigma}}~\![\delta g_{\mu\nu}]_{\textstyle {}_1} +
1\leftrightarrow 2 \;.
\end{equation}
From the defining property (\ref{57}) of the delta-pseudo-function
${\rm Pf}\Delta_1$, we can re-write (\ref{81}) in the equivalent form

\begin{equation}\label{82}
\delta I_{\rm particle} = \case{1}{2} m_1 c \int_{-\infty}^{+\infty} 
dt~\!{v_1^\mu v_1^\nu \over \sqrt{-[g_{\rho\sigma}]_{\textstyle {}_1} 
v_1^\rho v_1^\sigma}}<{\rm Pf}\Delta_1,\delta g_{\mu\nu}>+ 1\leftrightarrow 2 \;.
\end{equation}
Now, recall that the duality bracket is defined by the partie finie of
the three-dimensional integral [{\it cf} (\ref{9})], so the latter
expression can be cast into the standard form appropriate to the
definition of a stress-energy tensor $T^{\mu\nu}_{\rm particle}$,
namely

\begin{equation}\label{83}
\delta I_{\rm particle} = \case{1}{2} \int_{-\infty}^{+\infty} 
dt~\!<\sqrt{g}~\!T^{\mu\nu}_{\rm particle}~\!,\delta g_{\mu\nu}> \;.
\end{equation}
The only difference with the standard definition is that the partie
finie takes care of the divergencies at the positions of the
particles.  By comparing (\ref{82}) and (\ref{83}), we readily find
that the corresponding stress-energy tensor density is given by

\begin{equation}\label{84}
\sqrt{g}~\!T^{\mu\nu}_{\rm particle}=m_1 c ~\!{v_1^\mu v_1^\nu \over 
\sqrt{-[g_{\rho\sigma}]_{\textstyle {}_1} v_1^\rho v_1^\sigma}}
~\!{\rm Pf}\Delta_1 
+ 1\leftrightarrow 2 \;.
\end{equation}
The stress-energy tensor itself comes immediately from the rule of
multiplication of pseudo-functions (\ref{58}):

\begin{equation}\label{85}
T^{\mu\nu}_{\rm particle}=m_1 c ~\!{v_1^\mu v_1^\nu \over
\sqrt{-[g_{\rho\sigma}]_{\textstyle {}_1} v_1^\rho v_1^\sigma}}
~\!{\rm Pf}\left({\Delta_1\over \sqrt{g}}\right) + 1\leftrightarrow 2
\;,
\end{equation}
This tensor takes the same form as the stress-energy tensor of test
particles moving on a smooth background, except that the role of the
background field is now played by the metric generated by the
particles, regularized following the prescription (\ref{56}). Notice
in particular that the factor $1/\sqrt{g}$ inside the partie finie
sign ${\rm Pf}$ should not be replaced by its reguralized value at 1
[see (\ref{60})]. We propose the tensor (\ref{85}) as a model of
particles in the post-Newtonian approximation. From the product rules
for pseudo-functions, we get the matter source term in the right-hand
side of (\ref{78}) as

\begin{equation}\label{86}
g~\!T^{\mu\nu}_{\rm particle}=m_1 c ~\!{v_1^\mu v_1^\nu \over
\sqrt{-[g_{\rho\sigma}]_{\textstyle {}_1} v_1^\rho v_1^\sigma}}
~\!{\rm Pf}\left(\sqrt{g}\Delta_1\right) 
+ 1\leftrightarrow 2 \;.
\end{equation}
The post-Newtonian iteration of the field equations in
\cite{BF00,BFeom} is based on the latter expression of the matter source
term.

We now derive the equations of motion of the particle 1 from the
covariant conservation of the stress-energy tensor (\ref{85}):

\begin{equation}\label{87}
\nabla_\nu T^{\mu\nu}_{\rm particle}=0\;.
\end{equation}
Notice that thanks to the presence of the delta-pseudo-function, we
know that the derivative is ``ordinary'' and satisfies the Leibniz
rule in the sense of (\ref{60c}). Thus, we can transform $\nabla_\nu
T^{\mu\nu}_{\rm particle}$ in the standard way and find that the
equation (\ref{87}) is equivalent, like in the case of continuous
sources, to the alternative form

\begin{equation}\label{88}
\partial_\nu\left(\sqrt{g}~\! g_{\lambda\mu}~\!T^{\mu\nu}_{\rm particle}\right)
=\case{1}{2}\sqrt{g}~\!\partial_\lambda g_{\mu\nu}~\!T^{\mu\nu}_{\rm particle}\;.
\end{equation}
Then, we integrate (\ref{88}) over a closed volume ${\sc V}_1$
surrounding the particle 1 exclusively. The role of the
three-dimensional integral is played here by the duality bracket
defined by (\ref{9}). Let us denote by ${\bf 1}_{{\sc V}_1}$ the
characteristic function of the volume ${\sc V}_1$, such that ${\bf
1}_{{\sc V}_1}({\bf x})=1$ if ${\bf x}\in {\sc V}_1$ and ${\bf
1}_{{\sc V}_1}({\bf x})=0$ otherwise [notably, ${\bf 1}_{{\sc
V}_1}({\bf y}_2)=0$]. Thus, we consider

\begin{equation}\label{89}
<\partial_\nu\left(\sqrt{g}~\! g_{\lambda\mu}~\!T^{\mu\nu}_{\rm
particle}\right),{\bf 1}_{{\sc
V}_1}>=<\case{1}{2}\sqrt{g}~\!\partial_\lambda
g_{\mu\nu}~\!T^{\mu\nu}_{\rm particle}~\!,{\bf 1}_{{\sc V}_1}>\;.
\end{equation}
(Though ${\bf 1}_{{\sc V}_1}$ does not belong to the class ${\cal F}$,
it is locally integrable on ${\mathbb R}^3$ and we know that the
duality bracket applies on such functions as well; see \cite{BFreg}.)
The partial derivative $\partial_\nu$ in the left-hand side is split
into a time-derivative and a space-derivative.  Following the rule
(\ref{60a}), the spatial derivative $\partial_i$ is shifted to the
right side of the bracket, where it applies on the characteristic
function ${\bf 1}_{{\sc V}_1}$. Because of the presence of the
delta-pseudo-function, the derivative of ${\bf 1}_{{\sc V}_1}$ is to
be taken in an ordinary sense and is zero. Following the rule (9.7) in
\cite{BFreg}, an analogous reasoning is valid for the time-derivative
$\partial_0=\case{1}{c}\partial_t$ which can thus simply be put
outside the bracket. Thus, we get

\begin{equation}\label{90}
{d\over cdt}\biggl\{<\sqrt{g}~\! g_{\lambda\mu}~\!T^{\mu 0}_{\rm
particle}~\!,{\bf 1}_{{\sc
V}_1}>\biggr\}=<\case{1}{2}\sqrt{g}~\!\partial_\lambda
g_{\mu\nu}~\!T^{\mu\nu}_{\rm particle}~\!,{\bf 1}_{{\sc V}_1}>\;.
\end{equation}
Next, we insert into (\ref{90}) the specific expression (\ref{84}) of
the stress-energy density of particles. Because of the presence of the
function ${\bf 1}_{{\sc V}_1}$ only the part corresponding to the
particle 1 contributes, and we obtain

\begin{equation}\label{91}
{d\over dt}\Biggl\{{v_1^\mu \over \sqrt{-[g_{\rho\sigma}]_{\textstyle
{}_1} v_1^\rho v_1^\sigma}} <{\rm
Pf}\bigl(g_{\lambda\mu}\Delta_1\bigr),{\bf 1}_{{\sc
V}_1}>\Biggr\}=\case{1}{2}{v_1^\mu v_1^\nu \over
\sqrt{-[g_{\rho\sigma}]_{\textstyle {}_1} v_1^\rho v_1^\sigma}} <{\rm
Pf}\bigl(\partial_\lambda g_{\mu\nu}\Delta_1\bigr),{\bf 1}_{{\sc
V}_1}>\;.
\end{equation}
Finally, the effect of the brackets in both sides of the latter
equation is to take the value at the point 1 in the sense of the
Lorentzian regularization (\ref{56}). Thereby our final result reads
as

\begin{equation}\label{92}
{d\over dt}\Biggl({[g_{\lambda\mu}]_{\textstyle {}_1} v_1^\mu\over
\sqrt{-[g_{\rho\sigma}]_{\textstyle {}_1} v_1^\rho
v_1^\sigma}}\Biggr)=\case{1}{2}{[\partial_\lambda
g_{\mu\nu}]_{\textstyle {}_1}v_1^\mu v_1^\nu\over
\sqrt{-[g_{\rho\sigma}]_{\textstyle {}_1} v_1^\rho v_1^\sigma}}\;.
\end{equation}
The equations of motion of the particle 1 have the same formal
structure as the geodesic equations of a test particle. In separate
papers \cite{BF00,BFeom,ABF01} we use (\ref{92}) to derive
explicitely the equations of motion of the two particles at the 3PN
approximation.

\acknowledgments
We thank Misao Sasaki, Hideyuki Tagoshi and Takahiro Tanaka for
stimulating discussions, and for the permission to reproduce their
proof of the mathematical formula (\ref{a30}) at the end of the
appendix. The support of the JSPS short-term program for research in
Japan is gratefully acknowledged for a visit of one of us (L.B.) at
the University of Osaka during which this work was begun.

\appendix

\section{Solution of the equation (\ref{30})}

We are looking for the vector ${\bf z}_1$ satisfying the equation

\begin{equation}\label{a1}
{\bf z}_1={\bf y}_1\!\!\left(t-{1\over c^2}{\bf V}.({\bf x}-{\bf
z}_1)\right)\;,
\end{equation}
where ${\bf y}_1(t)$ represents a given smooth ($C^\infty$) time-like
trajectory and ${\bf V}$ a constant vector with norm $|{\bf
V}|<c$. Clearly, for a given trajectory, the solution ${\bf z}_1$
depends on the field point ${\bf x}$ as well as on time $t$. It was
shown in the text after (\ref{31}) that the application ${\bf
x}\rightarrow {\bf z}_1$ is contracting with fixed point ${\bf
y}_1$. Here, let us look for the solution ${\bf z}_1$ in the form of a
function of the coordinates,

\begin{equation}\label{a2}
{\bf z}_1={\bf z}_1({\bf x},t)\;.
\end{equation} 
From (\ref{a1}) we compute the partial derivatives of ${\bf z}_1$ with
respect to $t$ and $x^i$, considered to be independent, and readily
obtain

\begin{mathletters}\label{a3}\begin{eqnarray}
{\partial {\bf z}_1\over \partial x^i}&=&-{1\over c^2}\left[V_i-{\bf
V}.{\partial {\bf z}_1\over \partial x^i}\right] {\bf
v}_1\!\!\left(t-{1\over c^2}{\bf V}.({\bf x}-{\bf z}_1)\right)\\
{\partial {\bf z}_1\over \partial t}&=&\left[1+{1\over c^2}{\bf
V}.{\partial {\bf z}_1\over \partial t}\right] {\bf
v}_1\!\!\left(t-{1\over c^2}{\bf V}.({\bf x}-{\bf z}_1)\right)\;.
\end{eqnarray}\end{mathletters}$\!\!$
Contracting these equations with the vector ${\bf V}$ we can obtain
the scalar products ${\bf V}.{\partial {\bf z}_1\over \partial x^i}$
and ${\bf V}.{\partial {\bf z}_1\over \partial t}$, and use them back
into (\ref{a3}) with the result that

\begin{mathletters}\label{a4}\begin{eqnarray}
{\partial {\bf z}_1\over \partial x^i}&=&-{1\over c^2}V_i{{\bf
v}_1\over 1-{{\bf V}.{\bf v}_1\over c^2}}\\ {\partial {\bf z}_1\over
\partial t}&=&{{\bf v}_1\over 1-{{\bf V}.{\bf v}_1\over c^2}}\;,
\end{eqnarray}\end{mathletters}$\!\!$
where the velocity ${\bf v}_1$ is evaluated at the instant $t-{1\over
c^2}{\bf V}.({\bf x}-{\bf z}_1)$. In particular, we find that ${\bf
z}_1$ must be a solution of the following first-order differential
equation:

\begin{equation}\label{a5}
{\partial {\bf z}_1\over \partial x^i}=-{1\over c^2}V_i{\partial {\bf
z}_1\over \partial t}\;.
\end{equation}
Conversely, let us prove that a vector ${\bf z}_1$ that (i) satisfies
the differential equation (\ref{a5}) and (ii) admits ${\bf y}_1(t)$ as
a {\it fixed} point, i.e. is such that

\begin{equation}\label{a6}
{\bf z}_1\left({\bf y}_1(t),t\right)={\bf y}_1(t)\;,
\end{equation} 
necessarily satisfies the original equation (\ref{a1}). Such a ${\bf
z}_1({\bf x},t)$ being given, we perform in the equation (\ref{a5})
the change of variables $(x^i,t)\rightarrow (\rho_1^i,\tau_1)$ defined
by

\begin{mathletters}\label{a7}\begin{eqnarray}
\rho_1^i&=&x^i-z_1^i({\bf x},t)\;,\\
\tau_1&=&t-{1\over c^2}{\bf V}.\Bigl({\bf x}-{\bf z}_1({\bf x},t)\Bigr)\;.\label{a7b}
\end{eqnarray}\end{mathletters}$\!\!$
Using (\ref{a5}) it is easy to obtain the laws of transformation of
the partial derivatives:

\begin{mathletters}\label{a8}\begin{eqnarray}
{\partial\over \partial\rho^i_1}&=&{\partial\over \partial x^i}
+{1\over c^2}V_i{\partial\over \partial t}\;,\\ {\partial\over
\partial\tau_1}&=&{\partial\over \partial t}+B^i_{~j}{\partial
z_1^j\over \partial t}{\partial\over \partial x^i}\;,
\end{eqnarray}\end{mathletters}$\!\!$
where $B^i_{~j}$ denotes the matrix inverse of
$A^j_{~k}=\delta^j_k+{1\over c^2}V_k{\partial z_1^j\over \partial t}$
(i.e. $A^i_{~j}B^j_{~k}=\delta^i_k$; in the case considered here where
the velocities are strictly less than $c$ the matrix $A^i_{~j}$ is a
deformation of the unit matrix and thus admits an inverse). Now, under
the change of variables (\ref{a7}) the differential equation
(\ref{a5}) becomes simply

\begin{equation}\label{a9}
{\partial {\bf z}_1\over \partial \rho_1^i}={\bf 0}\;,
\end{equation}
whose general solution is an arbitrary function of the time variable
$\tau_1$. Therefore, there must exists a trajectory ${\bf Y}_1$ such
that

\begin{equation}\label{a10}
{\bf z}_1={\bf Y}_1(\tau_1)={\bf Y}_1\!\!\left(t-{1\over c^2}{\bf
V}.({\bf x}-{\bf z}_1)\right)\;.
\end{equation}
Imposing now that ${\bf y}_1(t)$ is a fixed point for this solution
${\bf z}_1$ in the sense of (\ref{a6}) leads immediately to

\begin{equation}\label{a11}
{\bf Y}_1(t)={\bf y}_1(t)\;,
\end{equation}
so the equation (\ref{a1}) is recovered exactly.  Thus, solving
(\ref{a1}) is equivalent to solving the differential equation
(\ref{a5}) supplemented by the condition (\ref{a6}). Notice that from
(\ref{a1}) or equivalently from (\ref{a5})-(\ref{a6}) we find that
${\bf z}_1$ tends to the fixed point in the ``non-relativistic'' limit
$c\to+\infty$, i.e.

\begin{equation}\label{a13}
\lim_{c\to +\infty}\{{\bf z}_1({\bf x},t)\}={\bf y}_1(t)\;.
\end{equation} 
This suggests to look for the solution ${\bf z}_1$ in the form of an
infinite series of relativistic corrections of successive orders
$1/c^{2n}$ [from (\ref{a5}) we know that ${\bf z}_1$ is a function of
$1/c^2$]. Thus, taking also into account the limit (\ref{a13}), we
pose

\begin{equation}\label{a14}
{\bf z}_1({\bf x},t) ={\bf y}_1(t)+\sum_{n=1}^{+\infty}{1\over
c^{2n}}\stackrel{n}{{\bf Z}}_1({\bf x},t)\;,
\end{equation} 
and we look for each one of the unknown coefficients
$\stackrel{n}{{\bf Z}}_1({\bf x},t)$. By replacing the series
(\ref{a14}) into both sides of the equation (\ref{a5}) and identifying
the factors of the powers of $1/c^2$ on each side we find, for any
$n\geq 1$,

\begin{equation}\label{a15}
{\partial \!\stackrel{n}{{\bf Z}}_1\over \partial x^i}=-V_i {\partial
\!\stackrel{n-1}{{\bf Z}_1}\over \partial t}\;,
\end{equation}
with the convention that $\stackrel{0}{{\bf Z}}_1={\bf y}_1(t)$. The
equations (\ref{a15}) are to be solved using the condition of fixed
point ${\bf y}_1$ [{\it cf} (\ref{a6})], which implies that, $\forall
n\geq 1$,

\begin{equation}\label{a16}
\stackrel{n}{{\bf Z}}_1({\bf y}_1(t),t)={\bf 0} \;.
\end{equation}
The solution of (\ref{a15})-(\ref{a16}) is found by induction over
$n$. As an induction hypothesis suppose that

\begin{equation}\label{a17}
\stackrel{n-1}{{\bf Z}_1}={(-)^{n-1}\over (n-1)!}\left({\partial\over 
\partial t}\right)^{\!n-2}\!\!\Bigl[({\bf V}.{\bf r}_1)^{n-1}{\bf v}_1\Bigr]\;,
\end{equation}
where ${\bf r}_1={\bf x}-{\bf y}_1$, and where the partial time
derivatives act on $t$ keeping the space coordinate ${\bf x}$ fixed:
for instance, $\partial {\bf r}_1/\partial t=-{\bf v}_1$ and $\partial
{\bf v}_1/\partial t=d{\bf v}_1/dt={\bf a}_1$, where ${\bf a}_1$ is
the acceleration.  Notice that (\ref{a17}) satisfies the condition
(\ref{a16}) because it involves $n-2$ partial time derivatives while
there is a factor $({\bf V}.{\bf r}_1)^{n-1}$ inside the brackets, so
after differentiation there will remain at least one factor ${\bf
V}.{\bf r}_1$ making the result be zero when ${\bf x}={\bf
y}_1$. Inserting (\ref{a17}) into the right-hand side of (\ref{a15})
we obtain the equation to be satisfied for the next-order coefficient,

\begin{equation}\label{a18}
{\partial \!\stackrel{n}{{\bf Z}}_1\over \partial x^i}=V_i
{(-)^{n}\over (n-1)!}\left({\partial\over \partial
t}\right)^{\!n-1}\!\!\Bigl[({\bf V}.{\bf r}_1)^{n-1}{\bf v}_1\Bigr]\;,
\end{equation}
which can be re-written equivalently in the form

\begin{equation}\label{a18'}
{\partial \!\stackrel{n}{{\bf Z}}_1\over \partial x^i}={\partial\over
\partial x^i}\Biggl\{{(-)^{n}\over n!}\left({\partial\over \partial
t}\right)^{\!n-1}\!\!\Bigl[({\bf V}.{\bf r}_1)^{n}{\bf
v}_1\Bigr]\Biggr\}\;,
\end{equation}
showing that the most general solution is necessarily of the type

\begin{equation}\label{a19}
\stackrel{n}{{\bf Z}}_1={(-)^{n}\over n!}\left({\partial\over 
\partial t}\right)^{\!n-1}\!\!\Bigl[({\bf V}.{\bf r}_1)^{n}{\bf v}_1\Bigr]
+{\bf C}(t)\;,
\end{equation}
where ${\bf C}(t)$ denotes an arbitrary vector depending only on time
$t$. However, this vector must be zero on account of the fact that the
result should be zero when ${\bf x}={\bf y}_1$. Therefore we have
proved by induction that

\begin{equation}\label{a19'}
\stackrel{n}{{\bf Z}}_1={(-)^{n}\over n!}\left({\partial\over 
\partial t}\right)^{\!n-1}\!\!\Bigl[({\bf V}.{\bf r}_1)^{n}{\bf v}_1\Bigr]\;,
\end{equation}
so the vector ${\bf z}_1$ solving at once (\ref{a5}) and (\ref{a6}),
or equivalently (\ref{a1}), takes the form of the rather interesting
infinite series

\begin{equation}\label{a20}
{\bf z}_1={\bf y}_1+\sum_{n=1}^{+\infty}{(-)^n\over c^{2n}
n!}\left(\partial\over \partial t\right)^{\!n-1}\!\!\Bigl[({\bf
V}.{\bf r}_1)^n {\bf v}_1\Bigr]\;,
\end{equation}
which constitutes the solution needed for our work in Section
\ref{III}.  Furthermore, subtracting ${\bf x}$ from this solution and
contracting with ${\bf V}$ we obtain after a short calculation the
quantity $\tau_1$ which was defined in (\ref{a7b}):

\begin{equation}\label{a21}
\tau_1=t+\sum_{n=1}^{+\infty}{(-)^n\over c^{2n} n!}\left(\partial
\over \partial t\right)^{\!n-1}\!\!\Bigl[({\bf V}.{\bf r}_1)^n \Bigr]\;.
\end{equation}

Now, recall that the latter quantity $\tau_1$ is such that ${\bf
z}_1={\bf y}_1(\tau_1)$. Therefore, we see that we can find an
alternative expression of the vector ${\bf z}_1$ by inserting into
${\bf y}_1(\tau_1)$ the series expansion (\ref{a21}) found for
$\tau_1$. Using an infinite Taylor expansion we are led to

\begin{equation}\label{a24}
{\bf z}_1={\bf y}_1+\sum_{p=0}^{+\infty}{1\over (p+1)!}{d^p {\bf
v}_1\over d t^p}\left(\sum_{n=1}^{+\infty}{(-)^n\over c^{2n}
n!}\left(\partial\over \partial t\right)^{\!n-1}\!\!\Bigl[({\bf
V}.{\bf r}_1)^n \Bigr]\right)^{\!p+1}\;.
\end{equation}
Each of the terms is composed of $p+1$ sums; accordingly we introduce
$p+1$ summation indices $n_1$, $\dots$, $n_p$, $n_{p+1}$ so that

\begin{eqnarray}\label{a25'}
{\bf z}_1&=&{\bf y}_1+\sum_{p=0}^{+\infty}{1\over (p+1)!}
{d^p {\bf v}_1\over d t^p}\sum_{n_1=1}^{+\infty}\dots
\sum_{n_p=1}^{+\infty}\sum_{n_{p+1}=1}^{+\infty}{(-)^{n_1+\dots 
+n_{p+1}}\over c^{2(n_1+\dots +n_{p+1})}}\nonumber\\
&\times&\left(\partial\over \partial
t\right)^{\!n_1-1}\!\!\Biggl[{({\bf V}.{\bf r}_1)^{n_1}\over
n_1!}\Biggr]\dots \left(\partial\over \partial t\right)^{\!n_p-1}
\!\!\Biggl[{({\bf V}.{\bf r}_1)^{n_p}\over
n_p!}\Biggr]\left(\partial\over \partial
t\right)^{\!n_{p+1}-1}\!\!\Biggl[{({\bf V}.{\bf r}_1)^{n_{p+1}}\over
n_{p+1}!}\Biggr]\;.
\end{eqnarray}
Next we pose $k=n_1+\dots +n_p+n_{p+1}$, replace the index $n_{p+1}$
by $k$, and operate $p+1$ commutations of summations to arrive at

\begin{eqnarray}\label{a25}
{\bf z}_1&=&{\bf y}_1+\sum_{k=1}^{+\infty}{(-)^k\over
c^{2k}}\sum_{p=0}^{k-1} {1\over (p+1)!}{d^p {\bf v}_1\over d
t^p}\sum_{n_1=1}^{q_1}\dots
\sum_{n_p=1}^{q_p}\nonumber\\
&\times&\left(\partial\over \partial
t\right)^{\!n_1-1}\!\!\Biggl[{({\bf V}.{\bf r}_1)^{n_1}\over
n_1!}\Biggr]\dots \left(\partial\over \partial t\right)^{\!n_p-1}
\!\!\Biggl[{({\bf V}.{\bf r}_1)^{n_p}\over
n_p!}\Biggr]\left(\partial\over \partial
t\right)^{\!n_{p+1}-1}\!\!\Biggl[{({\bf V}.{\bf r}_1)^{n_{p+1}}\over
n_{p+1}!}\Biggr]\;,
\end{eqnarray}
in which $n_{p+1}=k-\sum_{i=1}^p n_i$ and
$q_j=1+\sum_{i=j}^{p+1}(n_i-1)$ (with $1\leq j\leq p$). We must
identify the latter complicated expression with the simpler form of
the vector ${\bf z}_1$ given by (\ref{a20}). From identifying the
powers of $1/c^2$ in both expressions we immediately obtain

\begin{eqnarray}\label{a27}
&&\left(\partial\over \partial t\right)^{\!k-1} \!\!\Biggl[{({\bf
V}.{\bf r}_1)^{k}\over k!}{\bf v}_1\Biggr]=\sum_{p=0}^{k-1}{1\over
(p+1)!}{d^p {\bf v}_1\over d t^p}\sum_{n_1=1}^{q_1}\dots
\sum_{n_p=1}^{q_p}\nonumber\\
&&\qquad\times\left(\partial\over \partial t\right)^{\!n_1-1}
\!\!\Biggl[{({\bf V}.{\bf r}_1)^{n_1}\over n_1!}\Biggr]\dots
\left(\partial\over \partial t\right)^{\!n_p-1} \!\!\Biggl[{({\bf
V}.{\bf r}_1)^{n_p}\over n_p!}\Biggr]\left(\partial\over \partial
t\right)^{\!n_{p+1}-1} \!\!\Biggl[{({\bf V}.{\bf r}_1)^{n_{p+1}}\over
n_{p+1}!}\Biggr]\;.
\end{eqnarray}
Finally, from using the binomial formula for the derivative of a
product, we can identify in each side of the latter equation the
coefficients of each $d^p {\bf v}_1/d t^p$, and we arrive at the
relation, valid for any $p$ and any $k\geq p+1$,

\begin{eqnarray}\label{a28}
\sum_{n_1=1}^{q_1}\dots
\sum_{n_p=1}^{q_p}\left(\partial\over \partial t\right)^{\!n_1-1} \!\!
\Biggl[{({\bf V}.{\bf r}_1)^{n_1}\over n_1!}\Biggr]&\dots& \left(\partial
\over \partial t\right)^{\!n_p-1} \!\!\Biggl[{({\bf V}.{\bf r}_1)^{n_p}\over n_p!}
\Biggr]\left(\partial\over \partial t\right)^{\!n_{p+1}-1} \!\!\Biggl[{({\bf V}.
{\bf r}_1)^{n_{p+1}}\over n_{p+1}!}\Biggr]\nonumber\\
&=&{(p+1)(k-1)!\over (k-1-p)!}\left(\partial\over \partial
t\right)^{\!k-p-1} \!\!\Biggl[{({\bf V}.{\bf r}_1)^{k}\over
k!}\Biggr]\;.
\end{eqnarray}
The latter relation actually represents a quite general mathematical
formula because we have specified nothing about the scalar product
${\bf V}.{\bf r}_1$. Therefore, the relation (\ref{a28}) holds in fact
in the case of an arbitrary sufficiently differentiable function
$f(t)$, so

\begin{eqnarray}\label{a29}
\sum_{n_1=1}^{q_1}\dots
\sum_{n_p=1}^{q_p}\left(d\over d t\right)^{\!n_1-1} \!\!\Biggl[{f^{n_1}\over n_1!}
\Biggr]&\dots& \left(d\over d t\right)^{\!n_p-1} \!\!\Biggl[{f^{n_p}\over n_p!}
\Biggr]\left(d\over d t\right)^{\!n_{p+1}-1} \!\!\Biggl[{f^{n_{p+1}}\over n_{p+1}!}
\Biggr]\nonumber\\
&=&{(p+1)(k-1)!\over (k-1-p)!}\left(d\over d t\right)^{\!k-p-1}
\!\!\Biggl[{f^{k}\over k!}\Biggr]\;.
\end{eqnarray}

The equivalence obtained above between the formula (\ref{a1}) and the
differential equation (\ref{a5}) together with the auxiliary condition
(\ref{a6}) shows {\it indirectly} that the mathematical formula
(\ref{a29}) is correct. However, a {\it direct} proof of this formula
has been found by Tanaka, Sasaki and Tagoshi. Here we reproduce their
proof in the particular case where $p=1$, so that $q_1=k-1$ and
$n_2=k-n$ (where $n\equiv n_1$), in which case the formula reads, for
any $k\geq 2$,

\begin{equation}\label{a30}
\sum_{n=1}^{k-1}\left(d\over d t\right)^{\!n-1} \!\!\Biggl[{f^{n}\over n!}\Biggr]
\left(d\over d t\right)^{\!k-n-1} \!\!\Biggl[{f^{k-n}\over (k-n)!}\Biggr]=2(k-1)
\left(d\over d t\right)^{\!k-2} \!\!\Biggl[{f^{k}\over k!}\Biggr]\;.
\end{equation}
We replace $f(t)$ in (\ref{a30}) by its Fourier transform,
$f(t)=\int_{-\infty}^{+\infty}\case{d\omega}{2\pi}e^{i\omega t}{\tilde
f}(\omega)$, and readily find that in order to prove the formula
(\ref{a30}) it suffices to prove the statement that the equation

\begin{eqnarray}\label{a31}
\sum_{n=1}^{k-1}\left({k\atop n}\right)\Big(\omega_{(1}+\omega_2+\dots 
+\omega_n\Big)^{n-1}\Big(\omega_{n+1}+&\dots& +\omega_{k)}\Big)^{k-n-1}\nonumber\\
&=& 2(k-1)\Big(\omega_{1}+\omega_2+\dots +\omega_k\Big)^{k-2}
\end{eqnarray}
holds identically for any family of real frequencies $\omega_1$,
$\omega_2$, $\dots$, $\omega_k$. Most importantly, the parenthesis
around indices in the left side of (\ref{a31}) indicate the complete
symmetrization over the $k$ frequencies $\omega_1$, $\dots$,
$\omega_k$ [in addition, $\left({k\atop n}\right)$ denotes the
binomial coefficient]. Let us single out one of the frequencies, for
instance $\omega_k$, and re-write (\ref{a31}) in a form involving an
explicit symmetrization over the other $k-1$ frequencies, $\omega_1$,
$\dots$, $\omega_{k-1}$, only:

\begin{eqnarray}\label{a32}
\sum_{n=1}^{k-1}\left({k-1\atop n}\right)\Big(\omega_{(1}+\dots 
+\omega_n\Big)^{n-1}\Big(\omega_{n+1}+&\dots& +\omega_{k-1)}
+\omega_k\Big)^{k-n-1}\nonumber\\
&=& (k-1)\Big(\omega_{1}+\omega_2+\dots +\omega_k\Big)^{k-2}
\end{eqnarray}
(in which we have simplified a factor 2 in both sides of the
equation).  Furthermore, let us replace in the latter formula
$\omega_k$ by some sum $\omega_k+\dots +\omega_{k+s}$, and symmetrize
over the whole set of frequencies $\omega_1$, $\dots$,
$\omega_{k+s}$. This yields, for any $s$,

\begin{eqnarray}\label{a33}
\sum_{n=1}^{k-1}\left({k-1\atop n}\right)\Big(\omega_{(1}+\dots 
+\omega_n\Big)^{n-1}\Big(\omega_{n+1}+&\dots& +\omega_{k+s)}\Big)^{k-n-1}\nonumber\\
&=& (k-1)\Big(\omega_{1}+\omega_2+\dots +\omega_{k+s}\Big)^{k-2}\;.
\end{eqnarray}
Now we prove that the equation (\ref{a31}), or equivalently
(\ref{a32}), is true by induction on the integer $k$. Therefore, our
induction hypothesis is that (\ref{a32}) is correct for {\it any}
$k\leq K$, and from this we want to show that it is correct again for
$k=K+1$. Note that from our induction hypothesis we know that
(\ref{a33}) is also correct for any $k\leq K$ and {\it any}
$s$. Consider the sum defined by the left side of (\ref{a32}) in the
case where $k=K+1$, say

\begin{eqnarray}\label{a33'}
S_{K+1}=\sum_{n=1}^{K}\left({K\atop n}\right)\Big(\omega_{(1}+\dots
+\omega_n\Big)^{n-1}\Big(\omega_{n+1}+\dots
+\omega_{K)}+\omega_{K+1}\Big)^{K-n}\;,
\end{eqnarray}
where we recall that one of the frequencies, i.e. $\omega_{K+1}$, is
``artificially'' singled out. However, $S_{K+1}$ is also given by half
the left-hand side of (\ref{a31}) and is symmetric in $\omega_1$,
$\dots$, $\omega_{K+1}$. We want to show that $S_{K+1}$ is equal to
the right-hand side of (\ref{a32}) with $k=K+1$. To this end, we
transform $S_{K+1}$ with the help of the binomial formula, and obtain
after a short calculation

\begin{equation}\label{a34}
S_{K+1}=\sum_{l=0}^{K-1}{\omega_{K+1}^l\over l!}{K!\over
(K-l)!}\sum_{n=1}^{K-l}
\left({K-l\atop n}\right)\Big(\omega_{(1}+\dots +\omega_n\Big)^{n-1}
\Big(\omega_{n+1}+\dots +\omega_{K)}\Big)^{K-n-l}\;.
\end{equation}
Now we have two sums over $l$ and $n$, and it is easy to recognize
that the second sum, over $n$, can be simplified as soon as $l\geq 1$
by means of (\ref{a33}) which is correct by induction under the
condition that $k\leq K$ and for any $s$. Posing $K-l=k-1$ and $k+s=K$
we see that this condition is realized if and only if $l\geq 1$. After
simplification we find

\begin{equation}\label{a35}
S_{K+1}=K\Big(\omega_{1}+\dots +\omega_{K+1}\Big)^{K-1}+\Psi_{K+1}
\left(\omega_{1}, ~\!\dots ,~\!\omega_{K}\right)\;,
\end{equation}
where the first term is the result we want to obtain, and where the
second term is a certain function of the frequencies $\omega_1$,
$\dots$, $\omega_{K}$ but which does {\it not} depend on
$\omega_{K+1}$. The expression of $\Psi_{K+1}$ is given for
completeness as

\begin{equation}\label{a36}
\Psi_{K+1}=\sum_{n=1}^{K}
\left({K\atop n}\right)\Big(\omega_{(1}+\dots +\omega_n\Big)^{n-1}\Big(\omega_{n+1}
+\dots +\omega_{K)}\Big)^{K-n}-K\Big(\omega_{1}+\dots +\omega_{K}\Big)^{K-1}\;.
\end{equation}
Now we use the fact that $S_{K+1}$ is actually fully symmetric with
respect to the $K+1$ frequencies $\omega_{1}$, $\dots$,
$\omega_{K+1}$. Therefore the function $\Psi_{K+1}$ must be a pure
constant, independent on any $\omega_n$. Furthermore, we know also
that $S_{K+1}$ is a homogeneous polynomial of degree $K-1$ in all the
$\omega_{1}$, $\dots$, $\omega_{K+1}$, so this constant must in fact
be zero: $\Psi_{K+1}=0$. Finally we are able to conclude on the
desired result,

\begin{eqnarray}\label{a37}
S_{K+1}=K\Big(\omega_{1}+&\dots& +\omega_{K+1}\Big)^{K-1}\;.
\end{eqnarray}
Incidentally, notice that the equality $\Psi_{K+1}=0$ is itself a
consequence of the same mathematical formula, since it follows from
setting $k=K+1$ and posing $\omega_{K+1}=0$ in (\ref{a32}).

\end{document}